\documentclass[10pt,a4paper]{article}

\usepackage{jheppub}

\usepackage{epsfig,epsf}
\usepackage{amsmath}
\usepackage{amsthm}
\usepackage{amsfonts}
\usepackage{amssymb}
\usepackage{dsfont}
\usepackage{epstopdf}
\usepackage{multirow}

\usepackage{marvosym}

\usepackage{slashed}

\usepackage[active]{srcltx}
\usepackage{wick}
\def\II{\hbox{{1}\kern-.25em\hbox{l}}}

\DeclareMathOperator{\Li}{Li}
\DeclareMathOperator{\PCt}{PCt}
\newcommand \vev [1] {\langle{#1}\rangle}
\newcommand \VEV [1] {\left\langle{#1}\right\rangle}
\title{{\Large \textnormal{DESY 16-012}}\\[2mm]
Two-loop conformal generators for leading-twist operators in QCD}

\author[a]{V. M. Braun}
\author[b,a]{A. N. Manashov}
\author[b]{S. Moch}
\author[a]{and M. Strohmaier}

\affiliation[a]{
   Institut f\"ur Theoretische Physik, Universit\"at
   Regensburg \\ D-93040 Regensburg, Germany}
\affiliation[b]{
   Institut f\"ur Theoretische Physik, Universit\"at Hamburg\\
   D-22761 Hamburg, Germany}
\emailAdd{vladimir.braun@ur.de}

\emailAdd{alexander.manashov@desy.de}

\emailAdd{sven-olaf.moch@desy.de}

\emailAdd{matthias.strohmaier@ur.de}

\abstract{
QCD evolution equations in minimal subtraction schemes have a hidden
symmetry:
One can construct three operators that commute with the evolution kernel
and form an $SL(2)$ algebra,
i.e. they satisfy (exactly) the $SL(2)$ commutation relations.
In this paper we find explicit expressions for these operators to
two-loop accuracy going over
to QCD in non-integer $d=4-2\epsilon$ space-time dimensions at the
intermediate stage.
In this way conformal symmetry of QCD is restored on quantum level at the
specially chosen (critical)
value of the coupling, and at the same time the theory is regularized
allowing one to use the standard
renormalization procedure for the relevant Feynman diagrams.
Quantum corrections to conformal generators in $d=4-2\epsilon$  effectively
correspond
to the conformal symmetry breaking in the physical theory in four dimensions
and the $SL(2)$ commutation relations lead to nontrivial constraints on
the renormalization group equations
for composite operators.
This approach is valid to all orders in perturbation theory and the result
includes automatically all terms
that can be identified as due to a nonvanishing QCD $\beta$-function (in
the physical theory in four dimensions).
Our result can be used to derive three-loop evolution
equations for flavor-nonsinglet
quark-antiquark operators including mixing with the operators containing
total derivatives.
These equations govern, e.g.,  the scale dependence of generalized hadron
parton distributions and
light-cone meson distribution amplitudes.
       }


\begin{document}

\maketitle

\newpage

%%%%%%%%%%%%%%%%%%%%%%%%%%%%%%%%%%%%%%%%%%%%%%%%%%%%%%%%%%%%%%%%%%%%%%%%%%%%%%%%%%%%%%%%%%
\section{Introduction}\label{Sec:Introduction}
%%%%%%%%%%%%%%%%%%%%%%%%%%%%%%%%%%%%%%%%%%%%%%%%%%%%%%%%%%%%%%%%%%%%%%%%%%%%%%%%%%%%%%%%%%

Scale dependence of physical observables in strong interactions involving a large momentum
transfer is governed by the renormalization group (RG) equations for the corresponding
(composite) operators.
They have to be calculated to a sufficiently high order in perturbation theory in order
to make the theory description fully quantitative.
The anomalous dimensions of the leading twist-two operators are known to
NNLO accuracy (three loops), and these results have been converted to the
state-of-the-art NNLO evolution equations \cite{Moch:2004pa,Vogt:2004mw} for parton distributions
that are used in modern description of inclusive reactions, e.g., at the LHC.

A remarkable progress in accelerator and detector technologies in the last decades
has made possible the study of hard exclusive reactions with identified particles in the final state. Such studies
have become a prominent part of the research program at all major existing and planned accelerator facilities.
The relevant nonperturbative input in such processes involves operator matrix elements
between states with different momenta, dubbed generalized parton distributions (GPDs),
or vacuum-to-hadron matrix elements related to light-front hadron wave functions at small
transverse separations, the distribution amplitudes (DAs).
The different momenta in the initial and the final state complicates the RG equations since
mixing with the operators involving total derivatives has to be taken into account.
The arising mixing matrix (for a given moment, or operator dimension) is triangular
so that the diagonal entries correspond to the anomalous dimensions that are known to NNLO
accuracy, but the nondiagonal contributions require a dedicated calculation.

A direct calculation in higher orders is quite challenging, however, it has been known for some time~\cite{Mueller:1991gd}
that conformal symmetry of the QCD Lagrangian allows one to restore nondiagonal entries
in the mixing matrix and, hence, full evolution kernels at given order of perturbation theory
from the calculation of the special conformal anomaly at one order less.
This result was used to calculate the complete two-loop mixing matrix for
twist-two operators in QCD~\cite{Mueller:1993hg,Mueller:1997ak,Belitsky:1997rh},
and derive the two-loop evolution kernels for the GPDs~\cite{Belitsky:1998vj,Belitsky:1999hf,Belitsky:1998gc}.

In Ref.~\cite{Braun:2013tva} we have suggested an alternative technique, the difference being that
instead of studying conformal symmetry \emph{breaking} in the physical
theory~\cite{Mueller:1993hg,Mueller:1997ak,Belitsky:1997rh} we make use of the
\emph{exact} conformal symmetry of a modified theory -- QCD in $d=4-2\epsilon$ dimensions at critical coupling.
Exact conformal symmetry allows one to use algebraic group-theory methods to resolve the constraints on the
operator mixing and also suggests the optimal representation for the results in terms
of light-ray operators. In this way a delicate procedure of the restoration of the evolution kernels
as functions of two variables, e.g. momentum fractions, from the results for local operators can be avoided.

Utility of this modified approach was illustrated in~\cite{Braun:2013tva} on several examples to two- and
three-loop accuracy for scalar theories, and in~\cite{Braun:2014vba} on the example of the
two-loop evolution equation for flavor-nonsinglet operators in QCD. The present work is the first step towards
the three-loop calculation in QCD. Our main result is the calculation of the two-loop contribution to the generator
of special conformal transformations for flavor-nonsiglet leading-twist operators in
QCD in non-integer $d=4-2\epsilon$ space-time dimensions at critical coupling.

The presentation is organized as follows. Sect.~2 is introductory. We explain there the general strategy of our approach and
introduce the necessary formalism and notations. Sect.~3 and related Appendices A,B contain a detailed analysis of the scale and special conformal
Ward Identities (WI) in $d$-dimensional QCD. The expression in Eq.~\eqref{S+loop} for the $\ell$-loop quantum correction to the generator of
special conformal transformations is the main outcome of this analysis. In Sect.~4 we explain some technical issues that one encounters in the
calculation. The results for separate Feynman diagrams are collected in Appendix~C. Sect.~5 contains our principal result: the two-loop expression for
the generator of special conformal transformations. The two-loop expression for the evolution kernel in the light-ray operator representation~\cite{Braun:2014vba}
is given as well. The final Sect.~6 contains a short summary and outlook.

%%%%%%%%%%%%%%%%%%%%%%%%%%%%%%%%%%%%%%%%%%%%%%%%%%%%%%%%%%%%%%%%%%%%%%%%%%%%%%%%%%%%%%%%%%%%%%%%%%%%%%%%%%%%%%%%%%%%%%
\section{Conformal QCD}
%%%%%%%%%%%%%%%%%%%%%%%%%%%%%%%%%%%%%%%%%%%%%%%%%%%%%%%%%%%%%%%%%%%%%%%%%%%%%%%%%%%%%%%%%%%%%%%%%%%%%%%%%%%%%%%%%%%%%%

%%%%%%%%%%%%%%%%%%%%%%%%%%%%%%%%%%%%%%%%%%%%%%%%%%%%%%%%%%%%%%%%%%%%%%%%%%%%%%%%%%%%%%%%%%%%%%%%%%%%%%%%%%%%%%%%%%%%%%
\subsection{QCD in $4-2\epsilon$ dimensions at the critical point}
%%%%%%%%%%%%%%%%%%%%%%%%%%%%%%%%%%%%%%%%%%%%%%%%%%%%%%%%%%%%%%%%%%%%%%%%%%%%%%%%%%%%%%%%%%%%%%%%%%%%%%%%%%%%%%%%%%%%%%

We consider QCD in the $d=4-2\epsilon$ Euclidean space. The action reads
\begin{align}\label{SQCD}
S=\int d^dx \Big\{\bar q \slashed{D} q+\frac14 F_{\mu \nu}^aF^{a,\mu\nu}
- \bar c^a\partial_\mu(D^\mu c)^a +\frac1{2\xi}(\partial_\mu A^{a,\mu})^2\Big\}\,,
\end{align}
where
$D_\mu={\partial_\mu}-ig_B A_\mu^a T^a$  with $T^a$ being the $SU(N)$ generators in the fundamental (adjoint)
representation for quarks (ghosts). The bare coupling constant is $g_B=g M^\epsilon$ where $M$ is the scale parameter,
and the strength tensor is defined as usual
% Here $D^\mu c=\partial^\mu c-ig_0[A^\mu,c]$
%
\begin{align}\label{}
F_{\mu\nu}^a=\partial_\mu A_\nu^a -\partial_\nu A_\mu^a +g_B f^{abc}A_\mu^b A_\nu^c\,.
\end{align}
The renormalized action is obtained from~(\ref{SQCD}) by the replacement
\begin{align}\label{Zfactors}
q\to Z_q q\,, && A\to Z_A A\,,&& c\to  Z_c c\,, && g\to Z_g g\,,&& \xi\to Z_\xi \xi\,,
\end{align}
where $Z_\xi=Z_A^2$ and the renormalization factors are defined using minimal subtraction
\begin{align}
 Z = 1 + \sum\limits_{j=1}^\infty \epsilon^{-j} \sum\limits_{k=j}^\infty z_{jk} \left(\frac{\alpha_s}{4\pi}\right)^k\,,
\qquad \alpha_s = \frac{g^2}{4\pi}\,.
\end{align}
where $z_{jk}$ are $\epsilon$-independent constants.
Note that we do not send $\epsilon\to 0$ in the action and the renormalized correlation functions so that
they explicitly depend on $\epsilon$.

Formally the theory has two charges --- $g$ and $\xi$. The corresponding $\beta$-functions are
\begin{align}\label{}
\beta_g(g) = M\frac{d g}{dM} = g\big(-\epsilon - \gamma_g\big)\,,
&&
\beta_\xi(\xi,g) = M\frac{d\xi}{dM} = -2\xi\gamma_A\,,
\end{align}
where
\begin{align}
\gamma_g=M\partial_M \ln Z_g = \beta_0 \left(\frac{\alpha_s}{4\pi}\right) + \beta_1\left(\frac{\alpha_s}{4\pi}\right) ^2 +
 \mathcal{O}(\alpha_s^3)\,,  %\beta_0 = \frac{11}{3} N_c - \frac23 N_f
\end{align}
with
\begin{align}
\beta_0 = \frac{11}{3} N_c - \frac23 N_f, &&
\beta_1=\frac23\left[ 17 N_c^2-5N_c N_f -3 C_F N_f\right]\,.
\end{align}
The anomalous dimensions of the fields $ \Phi=\{q,\bar q, A,c,\bar c \}$ are defined as
\begin{align}
\gamma_\Phi = M\partial_M \ln Z_\Phi=\big(\beta_g \partial_g +\beta_\xi\partial_\xi\big)\ln Z_\Phi \,.
\end{align}
They are known to a high order, $\mathcal{O}(\alpha_s^5)$ for the quark anomalous dimension~\cite{Baikov:2014qja}.

In what follows we also use a notation
\begin{align}
          a = \frac{\alpha_s}{4\pi}\,, \qquad\quad \beta(a) = 2a \big(-\epsilon - \gamma_g\big)\,.
\end{align}

For a sufficiently large number of flavors, $N_f$, one obtains $\beta_0<0$. Therefore, there exists
a special (critical) value of the coupling, $g=g_*(\epsilon)$ such that $\beta_g(g_*)=0 $,
alias $ \epsilon=-\gamma_g(a_*) $ or, equivalently,
\begin{align}
a_*(\epsilon)=\left(\frac{g_*(\epsilon)}{4\pi}\right)^2=-\frac{\epsilon}{\beta_0}-\left(\frac{\epsilon}{\beta_0}\right)^2\frac{\beta_1}{\beta_0}+O(\epsilon^3)\,.
\end{align}
The $\beta$-function associated with the gauge parameter $\xi$ vanishes identically in the Landau gauge $\xi=0$. As a
consequence Green functions of the quark and gluon fields in Landau gauge at critical coupling enjoy scale
invariance \cite{Banks:1981nn,Hasenfratz:1992jv,Ciuchini:1999cv}.

Scale invariance usually implies conformal invariance of the theory: It is  believed that ``physically reasonable''
scale invariant theories are also conformally invariant, see Ref.~\cite{Nakayama:2013is} for a discussion. In
non-gauge theories conformal invariance for the Green functions of basic fields can be checked in perturbative
expansions~\cite{Sarkar:1974xh,Derkachov:1993uw}. For local composite operators a proof of conformal invariance is
based on the analysis of pair counterterms for the product of the trace of energy-momentum tensor and local
operators~\cite{Vasilev:2004yr}. In gauge theories, including QCD, conformal invariance does not hold for the
correlators of basic fields and can be expected only for the Green functions of gauge-invariant operators. Extra
complications are due to mixing of gauge-invariant operators with BRST variations and equation-of-motion (EOM)
operators. We will discuss these issues briefly in what follows.

Renormalization ensures finiteness of the correlation functions of the basic fields that are encoded in the
QCD partition function. Correlation functions with an insertion of a composite
operator, $\mathcal{O}_k$, possess additional divergences that are removed by the operator renormalization,
\begin{align}
  [\mathcal{O}_k] &= \sum_j \mathbb{Z}_{kj} \mathcal{O}_j\,,
\end{align}
where the sum goes over all operators with the same quantum numbers that get mixed;
$\mathbb{Z}_{kj}$ are the renormalization factors that have a similar
expansion in inverse powers of $\epsilon$ as in Eq.~(\ref{Zfactors}).
Here and below we use square brackets to denote renormalized composite
operators (in a minimal subtraction scheme).

Renormalized operators satisfy a RG equation with the anomalous dimension matrix
(or evolution kernel, in a different representation)
$\mathbb{H}\sim (-M\partial_M \mathbb{Z})\mathbb{Z}^{-1}$ (up to field renormalization)
which has a perturbative expansion with the coefficients that in minimal subtraction schemes
do not depend on $\epsilon$ by construction.
As a consequence, the anomalous dimension matrices are exactly the
same for QCD in $d$ dimensions that we consider
at the intermediate stage, and physical QCD in integer dimensions that is our final goal.
Namely, if in $d$-dimensional QCD at the critical point
\begin{align}
 \Big( M\partial_M + \mathbb{H}(a_\ast)\Big) [\mathcal{O}] = 0\,,\qquad
\mathbb{H}(a_\ast) = a_\ast \mathbb{H}^{(1)} + a_\ast^2 \mathbb{H}^{(2)} + \ldots
\end{align}
then at $d=4$ for arbitrary coupling
\begin{align}
 \Big( M\partial_M + \beta(a)\partial_a + \mathbb{H}(a)\Big) [\mathcal{O}] = 0\,,\qquad
\mathbb{H}(a) = a \mathbb{H}^{(1)} + a^2 \mathbb{H}^{(2)} + \ldots
\end{align}
with \emph{the same} matrices $\mathbb{H}^{(k)}$. All what one has to do in going over to the four-dimensional
world is to reexpress consistently all occurrences of $\epsilon = (4-d)/2$ in terms of the critical
coupling $\epsilon = \beta_0 a_\ast + \ldots $ and replace $a_\ast \mapsto a$ in the resulting expressions.
The requirement of large $N_f$ for the existence of the critical point is not principal since,
staying within perturbation theory, the dependence on $N_f$ is polynomial.
%Hence physical $N_f$ values can be obtained by analytic continuation.
In this sense the above connection holds for an arbitrary number of flavors.

Conformal symmetry of QCD in $d$-dimensions at the critical point means that evolution equations in physical QCD
in minimal subtraction schemes to all orders in perturbation theory have a hidden symmetry:
One can construct three operators that commute with $\mathbb{H}$ and form an $SL(2)$ algebra,
%Two more operators exist which, together with $\mathbb{H}$, form an $SL(2)$ algebra,
i.e. they satisfy (exactly) the  $SL(2)$ commutation relations.
As we will see below, perturbative expansion of these commutation relations produces
a nested set of equations that allow one to determine the non-diagonal parts of the anomalous
dimension matrices with a relatively small effort. A digression to the $4-2\epsilon$ dimensional world, from this
point of view, is just a technical trick in order to obtain the explicit expression for one of these operators,
the generator of special conformal transformations. To avoid misunderstanding, we stress that QCD in $d=4$ dimensions is
certainly not a conformal theory. The symmetry that we are going to exploit is the symmetry of \emph{RG equations} in QCD
in a specially chosen regularization scheme based on dimensional regularization with minimal subtraction.
The whole construction becomes simpler and more transparent going over from local operators to the corresponding
generating functions that are usually referred to as light-ray operators. This representation is introduced in
the next section.

%%%%%%%%%%%%%%%%%%%%%%%%%%%%%%%%%%%%%%%%%%%%%%%%%%%%%%%%%%%%%%%%%%%%%%%%%%%%%%%%%%%%%%%%%%%%%%%%%%%%%%%%%%%%%%%%%%%%%%
\subsection{Leading-twist operators}
%%%%%%%%%%%%%%%%%%%%%%%%%%%%%%%%%%%%%%%%%%%%%%%%%%%%%%%%%%%%%%%%%%%%%%%%%%%%%%%%%%%%%%%%%%%%%%%%%%%%%%%%%%%%%%%%%%%%%%

Poincare symmetry of the theory is enhanced at the critical point  $a=a_*$,
$\beta(a_*)=0$ by the dilatation symmetry (scale invariance) and symmetry under space-time inversion.
The subject of this work are flavor-nonsinglet  twist-two (symmetric and traceless) operators
\begin{align}
   \mathcal{O}_N(x) &=
 \sum_{k+m=N}c_{k,m} \, \bar q(x) (\stackrel{\leftarrow}{D}\!\cdot n)^m \slashed{n} (n\cdot\! \stackrel{\rightarrow}{D})^k q(x)
\label{O_N}
\end{align}
where $q$ and $\bar q$ are quark (antiquark) field operators that we tacitly assume to be of different flavor,
$D_\mu$ is a covariant derivative, and $n^\mu$ is an auxiliary light-like vector, $n^2=0$.
Symmetry transformations that act nontrivially
on these operators form the so-called collinear $SL(2,\mathbb{R})$ subgroup of the full conformal group that
leaves the light-ray $x^\mu=z n^\mu $ invariant, see Ref.~\cite{Braun:2003rp} for a review.

Collinear conformal transformations are generated by translations along the light-ray
direction $n^\mu$, special conformal transformations in the alternative light-like direction $\bar n$,
$\bar n^2=0$, $(n\bar n)=1$, and the combination of the dilatation and rotation in the $(n,\bar n)$ plane
\begin{align}
\mathbf{L}_-=-i\mathbf{P}_n,\! &&\!\mathbf{L}_+=\frac12i\mathbf{K}_{\bar n},\! &&\!
\mathbf{L_0}=\frac{i}2\left(\mathbf{D}-\mathbf{M}_{n\bar n}\right).
\label{gene}
\end{align}
Explicit expressions for the generators of translations $\mathbf{P}_\mu$, dilatations $\mathbf{D}$, special conformal
transformations $\mathbf{K}_\mu$ and Lorentz rotations $\mathbf{M}_{\mu\nu}$ can be found, e.g., in Ref.~\cite{Braun:2003rp}.
Here and below we use a shorthand notation $\mathbf{P}_n = n^\mu \mathbf{P}_\mu$ etc.
The generators defined in this way satisfy standard $SL(2)$ commutation relations
\begin{align}
{}[\mathbf{L}_{\pm},\mathbf{L}_0]=\pm \mathbf{L}_{\pm}\,, &&
{}[\mathbf{L}_{+},\mathbf{L}_-]=-2\mathbf{L}_0\,.
\label{Lcom}
\end{align}

Local composite operators can be classified according to irreducible representations of the $SL(2)$
algebra. A (renormalized) operator $[\mathcal{O}_N](x)$ is called conformal if it transforms covariantly under
the special conformal transformation:
\begin{align}
i\big[\mathbf{K}^\mu,[\mathcal{O}_N](x)\big] &=
\biggl[2 x^\mu (x
\partial)-x^2\partial^\mu+2\Delta^\ast_N x^\mu
+2x^\nu \left(n^\mu \frac{\partial}{\partial n^\nu}- n_\nu \frac{\partial}{\partial n_\mu}
\right)\biggr][\mathcal{O}_N](x)\,.
\label{Kmu}
\end{align}
Here $\Delta^\ast_N$ is the scaling dimension of the operator (at the critical point):
\begin{align}
i\big[\mathbf{D},[\mathcal{O}_N](x)\big]=\big(x\partial_x+\Delta^\ast_N\big)[\mathcal{O}_N](x)\,.
\label{DD}
\end{align}
As a consequence of having definite scaling dimension, a conformal operator $[\mathcal{O}_{N}]$
satisfies the RG equation
\begin{align}\label{OM}
\Big(M{\partial_M}+\gamma^*_N\Big)[\mathcal{O}_{N}]=0\,,
\end{align}
where $\gamma_N^*$ is the anomalous dimension at the critical point,
$\gamma_N^*=\gamma_N(a_*)$. The scaling dimension is given by the sum of the canonical and
anomalous dimensions, $$\Delta^\ast_N=\Delta_N+\gamma_N^*.$$
For the operators under
consideration $\Delta_N=2\Delta_q+N$ where $\Delta_q=d/2-1/2$ is the canonical dimension
of the quark field.

In a conformal theory, the correlation function of conformal operators is annihilated by the
generator of special conformal transformations,
\begin{align}
 (K^{(x_1)}_\mu + \ldots + K^{(x_n)}_\mu) \big\langle [\mathcal{O}_{N_1}(x_1)]\ldots [\mathcal{O}_{N_n}(x_n)]\big\rangle =0\,,
\label{annihilate}
\end{align}
where we added the superscripts $K^{(x_k)}_\mu$ to indicate explicitly the argument $x\mapsto x_k$ in the operators
(\ref{Kmu}); it is assumed that all space points $x_k$ are different. Eq.~(\ref{annihilate}) follows from the
requirement that the correlation function does not change under inversion of the coordinates
and the simultaneous transformation of the operators, it can be taken as a working definition of what
is meant by conformal symmetry of QCD at the critical point.

Each conformal operator $[\mathcal{O}_N]$
generates an irreducible representation of the $SL(2)$ algebra (conformal tower),
consisting of local operators obtained by adding total derivatives:
\begin{align}
 \mathcal{O}_{Nk}=(n\partial)^k [\mathcal{O}_N(0)],\,\qquad k=0,1,\ldots
\end{align}
such that
\begin{align}\label{3L}
\delta_-\mathcal{O}_{Nk}=~&\big[\mathbf{L}_-,\mathcal{O}_{Nk}\big]=~-\mathcal{O}_{Nk+1}\,,
\notag\\
\delta_{0\phantom{i}}\mathcal{O}_{Nk}=~&\big[\mathbf{L}_0,\,\mathcal{O}_{Nk}\big]=~(j_N+k)\mathcal{O}_{Nk}\,,
\notag\\
\delta_+\mathcal{O}_{Nk}=~&\big[\mathbf{L}_+,\mathcal{O}_{Nk}\big]=~k(2j_N+k-1)\mathcal{O}_{Nk-1}\,,
\end{align}
with the operator $[\mathcal{O}_N]$ itself being the highest weight vector,
$\big[\mathbf{L}_+,[\mathcal{O}_{N}]\big]=0$. Here $j_N$ is the so-called conformal spin of
the operator --- the half-sum of its scaling dimension and spin
\begin{align}
j_N=\frac12(\Delta^\ast_N+ N + 1 ) = \Delta_q + N  + \frac12 + \frac12 \gamma_N^*.
\end{align}
All operators $\mathcal{O}_{Nk}$ in a conformal tower have, obviously, the same anomalous
dimension~$\gamma_N^\ast$.

%%%%%%%%%%%%%%%%%%%%%%%%%%%%%%%%%%%%%%%%%%%%%%%%%%%%%%%%%%%%%%%%%%%%%%%%%%%%%%%%%%%%%%%%%%%%%%%%%%%%%%%%%%%%%%%%%%%%%%
\subsection{Light-ray operators }
%%%%%%%%%%%%%%%%%%%%%%%%%%%%%%%%%%%%%%%%%%%%%%%%%%%%%%%%%%%%%%%%%%%%%%%%%%%%%%%%%%%%%%%%%%%%%%%%%%%%%%%%%%%%%%%%%%%%%%

A renormalized light-ray operator,
\begin{align}
 [\mathcal{O}](x;z_1,z_2) = Z \mathcal{O}(x;z_1,z_2) = Z \bar q(x+z_1n)\slashed{n} q(x+z_2n),
\end{align}
where the Wilson line is implied between the quark fields on the light-cone,
is defined as the generating function for renormalized local operators:
\begin{eqnarray}
 [\mathcal{O}](x;z_1,z_2) &\equiv& \sum_{m,k} \frac{z_1^m z_2^k}{m!k!}
\left[\bar q(x) (\stackrel{\leftarrow}{D}\!\cdot n)^m \slashed{n} (n\cdot\! \stackrel{\rightarrow}{D})^k q(x)\right].
%[(D_n^m\bar q)(x) \slashed{n} (D_n^k q)(x)].
\label{LRO}
\end{eqnarray}
Due to Poincare invariance in most situations one can use $x=0$ in the definition of the light-ray operator~(\ref{LRO})
without loss of generality;  we will often use a shorthand notation
$$\mathcal{O}(z_1,z_2) \equiv \mathcal{O}(0; z_1,z_2).$$
The renormalization factor $Z$ is an integral operator in $z_1,z_2$ which is given by a series in~$1/\epsilon$
\begin{align}\label{Zfactor}
Z=1+\sum_{k=0}^\infty  \frac1{\epsilon^k} Z_k(a)\,, && Z_k(a)=\sum_{\ell=k}^\infty a^\ell Z_{k}^{(\ell)}\,.
\end{align}
The RG equation for the light-ray operator $[\mathcal{O}]$ takes the form
\begin{align}\label{RGO}
\Big(M{\partial_M}+\beta(a)\partial_a +\mathbb{H}(a)\Big)[\mathcal{O}](x;z_1,z_2)=0\,,
\end{align}
where $\mathbb{H}$ is an integral operator (evolution kernel) acting on the light-cone coordinates of the fields. It
is related to the renormalization factor~\eqref{Zfactor} as follows
\begin{align}\label{H-Z}
 \mathbb{H}(a)=-M\frac{d}{d M}\mathbb{Z} \mathbb{Z}^{-1}=2\gamma_q(a)+2\sum_{\ell=1}^\infty  \ell\,a^\ell Z_1^{(\ell)}\,,
\end{align}
where $\mathbb{Z}=ZZ_q^{-2}$.  The evolution kernel can be written as~\cite{Balitsky:1987bk}
\begin{align}
 \mathbb{H}(a)[\mathcal{O}](z_1,z_2) = \int_0^1 \!d\alpha\int_0^1\! d\beta\, h(\alpha,\beta)\, [\mathcal{O}](z_{12}^\alpha,z_{21}^\beta)\,,
%  + 2\gamma_q [\mathcal{O}](z_1,z_2)\,.
\label{hkernel}
\end{align}
where $h(\alpha,\beta)$ is a certain weight function (evolution kernel).
Here and below we use the notation
\begin{align}
z_{12}^\alpha =  z_1\bar\alpha+z_2\alpha && \bar\alpha=1-\alpha\,.
\end{align}
In perturbation theory  $h(\alpha,\beta)$ is given by a series in the coupling constant
\begin{align}
h(\alpha,\beta)=a\, h^{(1)}(\alpha,\beta)+a^2 h^{(2)}(\alpha,\beta)+\ldots
\end{align}
It is important to note that the fixed-order kernels $h^{(k)}(\alpha,\beta)$ in the $\overline{\text{MS}}$ scheme do not
depend on the space-time dimension by construction. Thus the dependence of $h(\alpha,\beta)$ on
$\epsilon$ in QCD in $4-2\epsilon$ dimensions at the critical point $a_* = a_*(\epsilon)$ comes exclusively  through the coupling constant.

Going over from the description in terms of conformal towers of local operators to the
light-ray operators essentially corresponds to going over to a different realization of conformal symmetry.
The light-ray operator $[\mathcal{O}(x;z_1,z_2)]$ can be expanded in terms of local operators~$\mathcal{O}_{Nk}$
\begin{align}\label{nlo2}
[\mathcal{O}(x; z_1,z_2)]=\sum_{Nk}\Psi_{Nk}(z_1,z_2)\,[\mathcal{O}_{Nk}(x)]\,,
\end{align}
where $\Psi_{Nk}(z_1,z_2)$ are homogeneous polynomials of degree $N+k$
\begin{align}
  \left(z_1\partial_{z_1} + z_2\partial_{z_2} -N-k\right)\Psi_{Nk}(z_1,z_2) =0\,
\end{align}
that we will refer to as coefficient functions.
The action of the generators of conformal transformations $\mathbf{L}_{\pm,0}$  on the light-ray operator is defined
via their relation to local operators,
\begin{align}
i\big[\mathbf{L}_{\boldsymbol{\alpha}},[\mathcal{O}](x=0;z_1,z_2)\big]  & =
 \sum_{Nk} \Psi_{Nk}(z_1,z_2)\, i\big[\mathbf{L}_{\boldsymbol{\alpha}},[\mathcal{O}_{Nk}(0)]\big]
\notag\\
& =\sum_{Nk} \sum_{m} \ell\,^{Nkm}_{\boldsymbol{\alpha}} \Psi_{Nm}(z_1,z_2)\,[\mathcal{O}_{Nm}(0)]\,,
\end{align}
where $\boldsymbol{\alpha}= \pm, 0$ and the coefficients $\ell\,^{Nkm}_{\boldsymbol{\alpha}}$ can be read off Eq.~(\ref{3L}).%
\footnote{For $x\slashed{=}0$ there are additional terms, cf.~(\ref{Kmu}).}
In this way the action of the generators $\mathbf{L}_{\pm,0}$ on the quantum fields in the light-ray operator
can be traded for the operators $S_{\pm,0}$ acting on the coefficient functions $\Psi_{Nk}(z_1,z_2)$:
\begin{align}
 \delta_{\pm,0} \Psi_{Nk}(z_1,z_2) &=  S_{\pm,0} \Psi_{Nk}(z_1,z_2)\,,
\end{align}
where
\begin{align}
   S_- \Psi_{Nk}(z_1,z_2) =& - \Psi_{Nk-1}(z_1,z_2)\,,
\notag\\
   S_0\, \Psi_{Nk}(z_1,z_2) =& (j_N+k) \Psi_{Nk}(z_1,z_2)\,,
\notag\\
   S_+ \Psi_{Nk}(z_1,z_2) =&  (k+1)(2j_N+k)\Psi_{Nk+1}(z_1,z_2)\,,
\label{steps}
\end{align}
and they can be represented as certain integro-differential operators $S_{\pm,0}$ acting on the quark coordinates
in the light-ray operator itself, in particular
\begin{align}
\big[ \mathbf{L}_+,[\mathcal{O}](x=0,z_1,z_2)\big] &=
\frac{i}{2}\big[ \bar n \mathbf{K},[\mathcal{O}](x=0,z_1,z_2)\big] ~=~  S_+ \big[\mathcal{O}\big](x=0,z_1,z_2)\,.
\end{align}
The generators $S_{\pm,0}$ in this (position-space) representation obey the  $SL(2)$ commutation relations
\begin{align}\label{sl2-comm}
{}[S_0,S_{\pm}]=\pm S_{\pm}\,, && {}[S_{+},S_-]= 2S_0\,,
\end{align}
(note a different sign as compared to the algebra of quantum operators (\ref{Lcom})),
and commute with the evolution kernel
\begin{align}\label{S-H}
[S_\alpha,\mathbb{H}]=0\,.
\end{align}
Explicit expressions for the generators in the interacting theory (at the critical point) are nontrivial
as, with the exception of $S_{-}$, they are modified by quantum corrections. One can write them
in the following form:
\begin{align}
   S_- &= S_-^{(0)}\,,
\notag\\
   S_0\, &= S_0^{(0)} +\Delta S_0 ~=~ S_0^{(0)}  -\epsilon+\frac12 \mathbb{H}(a_*)\,,
\notag\\
   S_+ &=  S_+^{(0)} + \Delta S_+ ~=~  S_+^{(0)} + (z_1+z_2)\Big(-\epsilon+ \frac12  \mathbb{H}(a_*)\Big)
+   (z_1-z_2)\,\Delta_+(a_*)\,,
\label{exactS}
\end{align}
where $S_\alpha^{(0)}$ are the canonical generators
\begin{align}\label{canon}
S^{(0)}_-&=-\partial_{z_1}-\partial_{z_2}\,,
\notag\\
S^{(0)}_0&=z_1\partial_{z_1}+z_2\partial_{z_2}+2,
\notag\\
S^{(0)}_+&=z_1^2\partial_{z_1}+z_2^2\partial_{z_2}+2(z_1+z_2)\,.
\end{align}
Note that quantum corrections to $S_0$ are completely determined by the evolution
kernel $\mathbb{H}$, whereas the generator of special conformal transformation
along the $\bar n$ direction, $S_+$, contains an additional contribution $\Delta_+$ that can be calculated
order by order in perturbation theory,
\begin{align}
\Delta_+(a_*)=a_* \Delta_+^{(1)}+a_*^2\, \Delta_+^{(2)}+\ldots.
\end{align}
By construction, the evolution kernel $\mathbb{H}$ and the operator $\Delta_+$ commute with the canonical generator
$S_0^{(0)}$. (It follows from the fact that only operators of the same canonical dimension can mix under
renormalization.)  Also, obviously, $[S^{(0)}_-,\Delta_+]=[S^{(0)}_-,\mathbb{H}]= 0$.
At the same time the evolution kernel $\mathbb{H}$ does not commute with $S_+^{(0)}$.

%%%%%%%%%%%%%%%%%%%%%%%%%%%%%%%%%%%%%%%%%%%%%%%%%%%%%%%%%%%%%%%%%%%%%%%%%%%%%%%%%%%%%%%%%%%%%%%%%%%%%%%%%%%%%%%%%%%%%%
\subsection{Conformal constraints for the evolution equation}
%%%%%%%%%%%%%%%%%%%%%%%%%%%%%%%%%%%%%%%%%%%%%%%%%%%%%%%%%%%%%%%%%%%%%%%%%%%%%%%%%%%%%%%%%%%%%%%%%%%%%%%%%%%%%%%%%%%%%%

One can show that the coefficient functions of the operators from the conformal tower are eigenfunctions
of the evolution kernel for the light-ray operator
\begin{align}
   \mathbb{H}(a_\ast)\,  \Psi_{Nk}(z_1,z_2) &= \gamma^\ast_N \, \Psi_{Nk}(z_1,z_2)
\end{align}
and have the following form%~\cite{Braun:2013tva}
\begin{align}
\Psi_{Nk}(z_1,z_2)=\varkappa_{Nk}  (S_+)^{k} z_{12}^N\,, && \varkappa_{Nk}=\frac{\Gamma(2j_N)}{k!\Gamma(2j_N+k)}\,,
\end{align}
Thus, the coefficient function of the conformal operator $[\mathcal{O}_{N}]$ is $\sim z_{12}^N$, and the coefficient
functions of the operators with extra total derivatives, $(n\partial)^k[\mathcal{O}_{N}]$, are obtained by the
repeated application of the ``step-up'' operator $S_+$.
As a consequence, anomalous dimensions of local operators correspond to the moments of the evolution
kernel for the light-ray operator
\begin{align}
 \gamma_N^\ast &= \int_0^1 \!d\alpha\int_0^1\! d\beta\, (1-\alpha-\beta)^N h(\alpha,\beta)\,.
\label{hmoments}
\end{align}

Using the representation for the generators in (\ref{exactS}) and expanding the commutation relation
$[S_+,\mathbb{H}]=0$ in a power series in the critical coupling $a_*$ one obtains a nested set of
equations~\cite{Braun:2013tva}
\begin{subequations}\label{nest}
\begin{align}
{}[S_+^{(0)},\mathbb{H}^{(1)}] &=~0\,,
\label{nest0}\\
{}[S_+^{(0)},\mathbb{H}^{(2)}] &=~[\mathbb{H}^{(1)},\Delta S_+^{(1)}]\,,
\label{nest1}\\
{}[S_+^{(0)},\mathbb{H}^{(3)}] &=~[\mathbb{H}^{(1)},\Delta S_+^{(2)}]+[\mathbb{H}^{(2)},\Delta S_+^{(1)}]\,.
\label{nest2}
\end{align}
\end{subequations}
The first equation (\ref{nest0})
expresses the usual wisdom that one-loop QCD evolution equations (in four dimensions) respect
conformal symmetry of the QCD Lagrangian~\cite{Makeenko:1980bh}.
In this case it can be shown that the corresponding kernel $h^{(1)}(\alpha,\beta)$
(up to trivial terms $\sim\delta(\alpha)\delta(\beta)$ that correspond to the unit operator)
takes the form~\cite{Braun:1999te}
\begin{align}\label{hinv}
h^{(1)} (\alpha,\beta) = \bar h (\tau)\,, && \tau = \frac{\alpha\beta}{\bar\alpha\bar\beta}\,
\end{align}
and is effectively a function of one variable $\tau$ called the conformal ratio.
This function can easily be reconstructed from its moments (\ref{hmoments}), alias from the anomalous dimensions.

This prediction is confirmed by explicit calculation~\cite{Balitsky:1987bk}:
\begin{align}
\mathbb{H}^{(1)}f(z_1,z_2)&=4C_F\biggl\{
\int_0^1d\alpha\frac{\bar\alpha}{\alpha}\Big[2f(z_1,z_2)-f(z_{12}^\alpha,z_2)-f(z_1,z_{21}^\beta)\Big]
\notag\\
&\quad
-\int_0^1d\alpha\int_0^{\bar\alpha}d\beta \, f(z_{12}^\alpha,z_{21}^\beta)
+\frac12 f(z_1,z_2)
\biggr\}\,.
\label{Honeloop}
\end{align}
The corresponding one-loop kernel $h^{(1)}(\alpha,\beta)$ can be written in the following,
remarkably simple form~\cite{Braun:1999te}
\begin{align}
      h^{(1)}(\alpha,\beta) = -4 C_F\left[\delta_+(\tau) + \theta(1-\tau)-\frac12\delta(\alpha)\delta(\beta)\right],
\label{QCD-LO}
\end{align}
where the regularized $\delta$-function, $\delta_+(\tau)$, is defined as
\begin{align}
\int d\alpha d\beta\, \delta_+(\tau)f(z_{12}^\alpha,z_{21}^\beta)&\equiv\int_0^1 d\alpha\int_0^{1} d\beta\, \delta(\tau)
\Big[f(z_{12}^\alpha,z_{21}^\beta)-f(z_1,z_2)\Big]
\notag\\
&=-
\int_0^1 d\alpha \frac{\bar \alpha}{\alpha}\Big[2f(z_1,z_2)-f(z_{12}^\alpha,z_2)-f(z_1,z_{21}^{\alpha})\Big].
\label{delta-function}
\end{align}
Taking appropriate matrix elements and making a Fourier transformation to the momentum fraction
space one can check that the expression in Eq.~(\ref{QCD-LO}) reproduces all classical
leading-order (LO) QCD evolution equations: the DGLAP equation for parton distributions,
the ERBL equation for the meson light-cone DAs, and the general evolution equation for GPDs.

The second equation (\ref{nest1}) states that breaking of conformal symmetry in the two-loop evolution kernel in the
usual sense, $[S_+^{(0)},\mathbb{H}^{(2)}] \slashed{=} 0$, is given by the commutator of the one-loop kernel and the
one-loop modification of the generator of special conformal transformation~\cite{Belitsky:1998gc,Braun:2014vba}
\begin{align}
\Delta S_+^{(1)} &= (z_1+z_2)\Big(\beta_0 + \frac12 \mathbb{H}^{(1)}\Big) + (z_1-z_2) \Delta^{(1)}_+\,,
\notag\\
\Delta^{(1)}_+[\mathcal{O}](z_1,z_2)
&= -2C_F\int_0^1d\alpha\Big(\frac{\bar\alpha}\alpha+\ln\alpha\Big)
\Big[[\mathcal{O}](z_{12}^\alpha,z_2)-[\mathcal{O}](z_1,z_{21}^\alpha)\Big]\,.
\label{Delta+1}
\end{align}
Since the canonical generator $S_+^{(0)}$  is nothing but the first-order differential operator,  Eq.~(\ref{nest1})
can be viewed as the first-order inhomogeneous differential equation for the two-loop kernel $\mathbb{H}^{(2)}$. The
general solution of this equation can be found as a special solution of the inhomogeneous equation, corresponding to
the symmetry breaking part of the evolution kernel, complemented by a general solution of the corresponding
homogeneous equation  $[S_+^{(0)},\mathbb{H}^{(2)}] = 0$ which has to be fixed by the requirement that the moments
(\ref{hmoments}) reproduce the known two-loop anomalous dimensions, see Ref.~\cite{Braun:2014vba} for the details.
The explicit expression for the two-loop kernel $h^{(2)}(\alpha,\beta)$ is given in Appendix B. It is equivalent to
the result for the two-loop splitting functions (in a different representation) for flavor-nonsinglet GPDs derived
in~\cite{Belitsky:1998gc} by a somewhat different method.

It is easy to see that this hierarchy continues to all orders in perturbation theory:
the evolution kernels at a given order of perturbation theory can be obtained from the
spectrum of anomalous dimensions at the same order and an additional calculation of the modification
of the generator of special conformal transformations at one order less.
In particular the three-loop evolution kernels require the knowledge of $S_+$ to two-loop accuracy,
see Eq.(\ref{nest2}). The corresponding calculation is the subject of this paper.

%%%%%%%%%%%%%%%%%%%%%%%%%%%%%%%%%%%%%%%%%%%%%%%%%%%%%%%%%%%%%%%%%%%%%%%%%%%%%%%%%%%%%%%%%%%%%%%%%%%%%%%%%%%%%%%%%%%%%%
\section{Scale and Conformal Ward Identities}
%%%%%%%%%%%%%%%%%%%%%%%%%%%%%%%%%%%%%%%%%%%%%%%%%%%%%%%%%%%%%%%%%%%%%%%%%%%%%%%%%%%%%%%%%%%%%%%%%%%%%%%%%%%%%%%%%%%%%%

Ward Identities (WI) follow, in general, from invariance
of suitable correlations functions under the change of variables in their path-integral representation,
corresponding to a symmetry transformation. The standard choice is the
correlation function of the composite operator in question with the set of fundamental fields.
In gauge theories and in particular in QCD it is more convenient to consider for the same purpose
the correlation functions of light-ray operators, which are gauge-invariant.

As mentioned above, the operator $S_+$ in the light-ray operator representation is defined as
the generator of special conformal transformations in the $\bar n$ direction acting
on the light-ray operator aligned in the opposite $n$-direction and centered at the origin, $x=0$:
\begin{align}
i\big[ \bar n \mathbf{K},[\mathcal{O}^{(n)}](x=0,z_1,z_2)\big]&=2(n\bar n) S_+ \big[\mathcal{O}^{(n)}\big](x=0,z_1,z_2)\,.
\end{align}
(Here we display explicitly the dependence on the auxiliary vector $n$ in the definition of the light-ray operator).
On the other hand, taking instead the $n$-projection and for arbitrary $x$ such that $(x\cdot n)=0$ one gets
\begin{align}
i\big[  n \mathbf{K},[\mathcal{O}^{(n)}](x,z_1,z_2)\big] &=- x^2(n\partial_x)\big[\mathcal{O}^{(n)}\big](x,z_1,z_2)\,,
\end{align}
or, changing $n\to\bar n$,
\begin{align}
i\big[  \bar n \mathbf{K},[\mathcal{O}^{(\bar n)}](x,z_1,z_2)\big] &=-x^2(\bar n\partial_x)\big[\mathcal{O}^{(\bar n)}\big](x,z_1,z_2)\,.
\end{align}
Consider the correlation function of two light-ray operators $[\mathcal{O}]^{(n)}(0,z)$ and
$[\mathcal{O}]^{(\bar n)}(x,w)$ aligned in opposite light-like directions and separated
by a transverse distance $(x\cdot n) =(x\cdot \bar n) = 0$:
\begin{align}\label{FInt1}
\mathcal{G}(x;z,w)=\VEV{[\mathcal{O}^{(n)}](0,z)\,[\mathcal{O}^{(\bar n)}](x,w)}\,,
%=\mathcal{N}\int D\Phi e^{-S_R(\Phi)}[\mathcal{O}^{(n)}](0,z)\,[\mathcal{O}^{(\bar n)}](x,w)\,.
\end{align}
where we use a shorthand notation  $z=\{z_1,z_2\}$, $w=\{w_1,w_2\}$.
%$\mathcal{N}$ is a normalization factor,  $S_R(\Phi)$ is the renormalized QCD action, $\Phi=\{A,q,\bar q,
%c,\bar c\}$ and the functional integration goes over all fields.

Conformal invariance of QCD at the critical point implies the constraint, cf.~(\ref{annihilate}),
\begin{eqnarray}
\label{EQ}
&&\hspace*{-4cm}\frac{i}{2}\VEV{[\bar n \mathbf{K},[\mathcal{O}^{(n)}](0,z)]\,[\mathcal{O}^{(\bar n)}](x,w)
+ [\mathcal{O}^{(n)}](0,z)\, [\bar n \mathbf{K},[\mathcal{O}^{(\bar n)}](x,w)]}
~=
\nonumber\\
&=& \left[S_+^{(z)}-\frac12 x^2 (\bar n\partial_x)\right]\mathcal{G}(x;z,w)
~=~ 0\,,
\end{eqnarray}
where the superscript $S_+^{(z)}$ reminds that it is a differential operator acting on the $z_1,z_2$ coordinates.
The explicit expression for $S_+^{(z)}$ can be derived starting from the path-integral representation
\begin{align}\label{FInt}
\mathcal{G}(x;z,w) &=
\mathcal{N}\int D\Phi\, e^{-S_R(\Phi)}[\mathcal{O}^{(n)}](0,z)\,[\mathcal{O}^{(\bar n)}](x,w)\,.
\end{align}
Here $\mathcal{N}$ is a normalization factor,  $S_R(\Phi)$ is the renormalized QCD action, $\Phi=\{A,q,\bar q,
c,\bar c\}$ and the functional integration goes over all fields.

Let us make a change of variables in the path-integral
\begin{align}
&\Phi\mapsto \Phi+\delta_D \Phi\,,  \qquad\qquad
\delta_D \Phi=\big(x\partial_x+\Delta_\Phi\big) \Phi(x)\,,
\label{D}\\
&\Phi\mapsto \Phi+\delta_K^\mu \Phi\,, \qquad\qquad
\delta_K^\mu \Phi=\Big(2x_\mu(x\partial)-x^2\partial_\mu +2\Delta_\Phi x_\mu -2\Sigma_{\mu\nu} x^\nu\Big) \Phi(x)\,,
\label{K}
\end{align}
corresponding to the dilatation and special conformal transformations, respectively, see e.g.~Ref.~\cite{Braun:2003rp}.
$\Sigma_{\mu\nu}$ in (\ref{K}) is the generator of spin rotations,
\begin{align*}
\Sigma_{\mu\nu} c=\Sigma_{\mu\nu}\bar c=0\,, && \Sigma_{\mu\nu} q=\frac i2\sigma_{\mu\nu} q\,, && \Sigma_{\mu\nu} A_\alpha=
g_{\nu\alpha} A_\mu-g_{\mu\alpha} A_\nu
\end{align*}
and $\Delta_\Phi$ are the scaling dimensions of the QCD fundamental fields, which are conveniently chosen as follows~\cite{Belitsky:1998gc}:
\begin{align}
\label{scalingdimension}
\Delta_q=\frac32-\epsilon, && \Delta_A=1, &&\Delta_c=0\,, && \Delta_{\bar c}=2-\epsilon\,.
\end{align}
The choice $\Delta_A=1$ ensures that the nonabelian field strength tensor transforms covariantly
under conformal transformations
\begin{align}
\delta_K^\mu F_{\alpha\beta}=\Big[2x_\mu(x\partial)-x^2\partial_\mu +4 x_\mu -2\Sigma_{\mu\nu} x^\nu\Big] F_{\alpha\beta}\,,
\end{align}
and the rationale for $\Delta_c=0$ is that for this choice a covariant derivative of the ghost field $D_\rho c(x)$
transforms as a vector field of dimension one, i.e. in the same way as the gluon field $A_\rho$.

Invariance of the path-integral representation of the correlation function of two
light-ray operators $\mathcal{G}(x;z,w)$ under the change of variables implies the identity
\begin{align}\label{SCWI}
\VEV{\delta[\mathcal{O}^{(n)}](0,z)\,[\mathcal{O}^{(\bar n)}](x,w)}+\VEV{[\mathcal{O}^{(n)}](0,z)\,\delta [\mathcal{O}^{(\bar n)}](x,w)}
=\VEV{\delta S_R\, [\mathcal{O}^{(n)}](0,z)\,[\mathcal{O}^{(\bar n)}](x,w)},
\end{align}
where $\delta=\delta_D$ and $\delta=\delta_K=\bar n_\mu \delta^\mu_K$ for scale and conformal
transformations,~\eqref{D} and~\eqref{K},
respectively, and $\delta S_R$ is the corresponding variation of the QCD action
\begin{align}\label{DC}
\delta_D S_R  & =
\int d^dx\, \mathcal{N}(x)\,,
\notag\\
\delta_K^\mu S_R  & =
\int d^dx\,2x^\mu \Big(\mathcal{N}(x)-(d-2) \partial^\rho \mathcal{B}_\rho (x)\Big)\,,
\end{align}
where
\begin{align}\label{NBmu}
\mathcal{N}(x)&= 2\epsilon\, \mathcal{L}^{YM+gf}_R = 2 \epsilon\,\left(\frac14 Z_A^2 F^2+\frac1{2 \xi} (\partial A)^2\right)\,,
\notag\\
\mathcal{B}_\rho(x)& =Z_c^2\bar c D^\rho c-\frac1\xi A^\rho (\partial A)\,.
\end{align}
Note that the coefficient of $\partial^\rho\mathcal{B}_\rho(x)$ in the conformal variation
does not vanish for $\epsilon\to 0$.
Hence the QCD action is not invariant under conformal transformations even for integer $d=4$ dimensions.
The operator $\mathcal{B}_\mu(x)$ can, however, be written as a BRST variation of $ \bar c^a A_\mu^a$~\cite{Belitsky:1998gc}, see
Appendix.~\ref{app:BRST}. Thus this term does not contribute to correlation functions  of gauge-invariant
operators~\cite{Collins} and can be dropped in most cases, which greatly  simplifies  the analysis.

In what follows we analyze the structure of the Ward Identities~\eqref{SCWI} in detail.

%%%%%%%%%%%%%%%%%%%%%%%%%%%%%%%%%%%%%%%%%%%%%%%%%%%%%%%%%%%%%%%%%%%%%%%%%%%%%%%%%%%%%%%%%%%%%%%%%%%%%%%%%%%%%%%%%%%%%%%
\subsection{Scale Ward Identity}
%%%%%%%%%%%%%%%%%%%%%%%%%%%%%%%%%%%%%%%%%%%%%%%%%%%%%%%%%%%%%%%%%%%%%%%%%%%%%%%%%%%%%%%%%%%%%%%%%%%%%%%%%%%%%%%%%%%%%%%
Let us first consider the scale, or dilatation, WI (SWI). The variation of the renormalized operators on the
l.h.s. of Eq.~\eqref{SCWI} is given by
\begin{align}
\delta_D[\mathcal{O}^{(n)}](x,z)=Z\delta_D\mathcal{O}^{(n)}(x,z)=\Big(x\partial_x +\sum_{i=1,2}z_i\partial_{z_i} + 3-2\epsilon\Big)[\mathcal{O}^{(n)}](x,z),
\end{align}
where we used that the renormalization $Z$-factor commutes with the operator
$\sum_{i}z_i\partial_{z_i}$ counting the canonical dimension.
(This is nothing but a usual observation that only the operators of the same canonical dimension mix under renormalization.)
We obtain
\begin{align}\label{scaleWI}
\Big(x\partial_x+\sum_{i=1,2}z_i\partial_{z_i}+\sum_{i=1,2}w_i\partial_{w_i}+6-4\epsilon\Big)\,
 \mathcal{G}(x;z,w)=\VEV{\delta_D S_R\, [\mathcal{O}^{(n)}](0,z)\,[\mathcal{O}^{(\bar n)}](x,w)}\,.
\end{align}
Since the l.h.s. can also be written as a derivative over the scale parameter $M\partial_M\mathcal{G}(x;z,w)$
and the RG equations for the light-ray operators take the form \eqref{RGO}, the
expression on the r.h.s. of~\eqref{scaleWI} that contains the variation of the action $\delta_D S_R$~\eqref{DC}
can be written as
\begin{align}\label{SRE}
\VEV{\delta_D S_R\, [\mathcal{O}^{(n)}](0,z)\,[\mathcal{O}^{(\bar n)}](x,w)}
=-\left(\beta(a)\partial_a + \mathbb{H}^{(z)}(a)+\mathbb{H}^{(w)}(a)\right)\mathcal{G}(x;z,w)\,.
\end{align}
It is instructive to derive this result by a direct calculation using a method that can be generalized to the more
complicated case of the conformal WI (CWI) (see also Refs.~\cite{Belitsky:1998gc,Belitsky:1998vj}).

The starting observation is that correlation functions of the basic fields with an insertion of the
operator~$\mathcal{N}(x)$~\eqref{NBmu} are finite, as follows from the structure of the corresponding
scale and conformal WIs. Thus $\mathcal{N}(x)$ can be expanded in terms of \emph{renormalized} operators
and the coefficients in this expansion can be fixed (apart from certain terms involving total derivatives)
from the renormalization group analysis. The result
reads~\cite{Spiridonov:1984br,Belitsky:1998vj,Belitsky:1998gc,Braun:2003rp,Collins,Vasilev:2004yr}
\begin{align}\label{NER}
\mathcal{N}(y)  &=-\frac{\beta(a)}{a}\left[\mathcal{L}^{YM+gf}%+\mathcal{L}^{gf}%-\frac12 \Omega_A
\right]-(\gamma_A+\gamma_g)\Omega_A-
\sum_{\Phi\neq A} \gamma_\Phi \Omega_{\Phi}
+ \frac{\gamma_A}{\xi}[(\partial A)^2]
+z_c\partial^\mu \Omega_\mu +z_b \partial_\mu [\mathcal{B}^\mu]\,,
\end{align}
where $\Omega_\Phi$ is an EOM operator, $\Omega_\Phi=\Phi(y)\Big({\delta S_R}/{\delta\Phi(y)}\Big)$ and
$\partial^\mu \Omega_\mu=\Omega_{\bar c}-\Omega_c = \partial^\mu[\bar c {D}_\mu c - \partial_\mu\bar c\, c]$.
The constants $z_c(g,\xi)$ and $z_b(g,\xi)$ cannot be determined in this method.
In order to make the presentation self-contained we explain their derivation in Appendix~\ref{App:action}.

The last term in~\eqref{NER}, being a BRST variation, does not contribute to the correlation function in~\eqref{scaleWI}.
The ghost EOM terms $\Omega_{\bar c}$ and $\Omega_{c}$ also do not contribute since the light-ray operators
do not contain ghost fields, e.g.,
\begin{align*}
\int\! d^dy \VEV{\Omega_{\bar c}(y)[\mathcal{O}^{(n)}](0,z)[\mathcal{O}^{(\bar n)}](x,w)}=
-\int\! d^dy\,\Big\langle{\bar c}(y)\frac{\delta}{\delta \bar c(y)}\left([\mathcal{O}^{(n)}](0,z)
{}[\mathcal{O}^{(\bar n)}](x,w)\right)\!\!\Big\rangle=0\,.
\end{align*}
Further,  the gauge fixing term can be replaced by the sum of EOM terms using Eqs.~\eqref{N2} and \eqref{fixing-B},
\begin{align}
  \frac{\gamma_A}{\xi}[(\partial A)^2] \to -\gamma_A\sum_{\Phi=A,q,\bar q} \Omega_\Phi\, \xi\partial_\xi \ln Z_\Phi\,.
\label{eq:gaugefix}
\end{align}
Note that the coefficients $\xi\partial_\xi \ln Z_\Phi$ are given by a series in $1/\epsilon$ without a constant term.
Since the WI is finite, all such singular terms must cancel in the final answer and it is sufficient, in principle,
to trace the nonsingular terms only. In other words, although the terms in \eqref{eq:gaugefix} contribute to the WI,
their only role is to cancel some other singular contributions. It is instructive, nevertheless,  to
trace these cancellations explicitly.

Thus we can replace $\mathcal{N}(y)$ by a somewhat simpler expression
\begin{align}\label{Nt}
\widetilde{\mathcal{N}}(y) =-\frac{\beta(a)}{a}\left[\mathcal{L}^{YM+gf} \right]-(\gamma_A+\gamma_g)\Omega_A
-\gamma_q(\Omega_q+\Omega_{\bar q})
-2\gamma_A\sum_{\Phi=A,q,\bar q} \Omega_\Phi\, \xi\partial_\xi \ln Z_\Phi\,.
\end{align}
 The EOM contributions give rise to contact terms that can be evaluated integrating by parts in the path-integral
\begin{align}
\Big\langle{\Omega_\Phi(y)O_1\, O_2}\Big\rangle&=
\Big\langle \Phi(y)\frac{\delta O_1}{\delta\Phi(y)} O_2\Big\rangle+
\Big\langle O_1\Phi(y)\frac{\delta O_2}{\delta\Phi(y)} \Big\rangle.
\end{align}
The quark and the antiquark EOM operators, $\Omega_{q\bar q}=\Omega_q+\Omega_{\bar q}$, give together
\begin{align}
\int d^dy\,\VEV{\Omega_{q\bar q}(y)[\mathcal{O}^{(n)}](0,z)\,[\mathcal{O}^{(\bar n)}](x,w)}=
4 \VEV{[\mathcal{O}^{(n)}](0,z)\,[\mathcal{O}^{(\bar n)}](x,w)},
\end{align}
so that
\begin{align}\label{SRE1}
\VEV{\delta_D S_R\, [\mathcal{O}^{(n)}](0,z)\,[\mathcal{O}^{(\bar n)}](x,w)}_{\Omega_{q\bar q}}
&=-4\big[\gamma_q + 2\gamma_A \xi\partial_\xi \ln Z_q\big] \mathcal{G}(x;z,w)\,.
\end{align}
Gluon EOM contributions are more complicated
(because light-ray operators contain terms with an arbitrary number of gluon fields), but we will show that they cancel.

The main contribution comes from the insertion of the renormalized Lagrangian
\begin{align}\label{SRE2}
\VEV{\delta_D S_R\, [\mathcal{O}^{(n)}](0,z)\,[\mathcal{O}^{(\bar n)}](x,w)}_{\mathcal{L}}
& = -\frac{\beta(a)}{a} \int \!d^d y
\VEV{ \left[\mathcal{L}^{YM+gf}\right](y) [\mathcal{O}^{(n)}](0,z)\,[\mathcal{O}^{(\bar n)}](x,w)}.
\end{align}
Since this correlation function involves three renormalized operators, the counterterms corresponding
to operator renormalization are already subtracted. All remaining divergences
correspond to pair counterterms for the contraction of $[\mathcal{L}^{YM+gf}](y)$ and one of the light-ray
operators, $y\to 0$ or $y\to x$.
We can write, schematically,
\begin{align}\label{A3}
\VEV{ \left[\mathcal{L}\right] [\mathcal{O}^{(n)}]\,[\mathcal{O}^{(\bar n)}]}
&=
\VEV{ \left[\mathcal{L} \mathcal{O}^{(n)} \mathcal{O}^{(\bar n)}\right]}
- \VEV{\PCt\Big(\mathcal{L} \mathcal{O}^{(n)}\Big)\, [\mathcal{O}^{(\bar n)}]}
- \VEV{ [\mathcal{O}^{(n)}]\,\PCt\Big(\mathcal{L} \mathcal{O}^{(\bar n)}\Big)}\,,
\end{align}
where $\PCt(A(x) B(y))$ denotes the pair counterterm for the contraction of the operators $A$ and $B$.
The first term on the r.h.s. of Eq.~\eqref{A3} is finite, by definition,
so that it does not contribute to \eqref{SRE2} at the critical point $\beta(a^*)=0$~\cite{Derkachov:1993uw,Vasilev:2004yr}.

Pair counterterms for the product of two arbitrary operators $A(x)$ and $B(y)$ have the following general structure~\cite{Vasilev:2004yr}

\begin{align}\label{CT}
\PCt\Big(A(x)B(y)\Big)=\delta(y-x) Z_i C_i(x)+ (\partial_y^\mu \delta(y-x)) \widetilde Z_i \widetilde C^i_\mu(x)+
\ldots\,,
\end{align}
where $C_i(x)$, $\widetilde C^i_\mu(x)$ are local operators and $Z_i$,  $\widetilde Z_i$ are singular coefficients.
The ellipses stand for the contributions with more than one derivative acting on the $\delta$-function.
For the case at hand only the terms without derivatives are relevant, which can be found without
explicit calculation.

To this end let us compare the structure of divergent contributions in the correlation function of the
two light-ray operators with and without the $\int d^dy [\mathcal L]$ insertion,
\begin{align}\label{singA}
  \sum_{D}\, \text{KR}'\,\left(D\Big\{\int d^d y\,\mathcal{L}^{YM+gh}(y)
  \mathcal{O}^{(n)}(x,z) \mathcal{O}^{(\bar n)}(x,w)\Big\}\right),
\end{align}
vs.
\begin{align}\label{singB}
 \sum_{D}\, \text{KR}'\,\left(D\Big\{ \mathcal{O}^{(n)}(x,z) \mathcal{O}^{(\bar n)}(x,w)\Big\}\right),
\end{align}
where the sum goes over all one-particle-irreducible (1PI) Feynman diagrams,
$D\{\ldots\}$ stands for the expression for a given diagram,
and the $\text{KR}'$ operation corresponds to taking singular part after subtraction of divergences in all
subgraphs.

An insertion of $\int d^dy \, \mathcal L^{YM+gf} $ in a generic Feynman diagram for the correlation function
generates two types of contributions: the kinetic term gives rise to an insertion (of unity)
in gluon propagators, and the interaction terms correspond to a replacement of one of the ``usual''
QCD vertices (three-gluon or four-gluon) by the ``special'' vertex which is in fact identically the same as
the ``usual'' one. The only effect of these substitutions is an extra combinatorial factor:
e.g. the kinetic term can be inserted in any gluon line, thus the original diagram is effectively
multiplied by the number of gluon lines, and similar for the vertices. It is easy to convince oneself that
the combined effect of all insertions is a multiplication of the diagram by the number of loops (minus one,
because the leading-order diagram for the correlation function already contains a loop).

Divergent contributions to the correlation function of the light-ray operators \eqref{singB} obviously
correspond to their renormalization.
Note that a single light-ray operator contains contributions with arbitrary many gluon fields, $\ell=0,1,\ldots$
so that the renormalized light-ray operator takes the form, schematically
\begin{align}\label{LRA}
[\mathcal{O}^{(n)}(x,z)]= Z\sum_{\ell=0}^\infty (Z_g Z_A)^\ell \mathcal{O}^{(n)}_\ell(x,z,A)\,.
\end{align}
The pair counterterms of interest are given by the same set of diagrams that give rise to the above product of
renormalization factors, correcting
for their combinatorial factors. Note that a multiplication by the number of loops (in a particular divergent subgraph)
amounts to taking a derivative $a\partial_a$.
Hence we can write e.g. for the operator counterterm corresponding to the contraction of
$\mathcal{L}^{YM+gf}(y)$ and $\mathcal{O}^{(n)}(x,z)$ (cf. Ref.~\cite{Belitsky:1998gc})
\begin{eqnarray}\label{PCT}
\int d^d y\, \PCt\Big(\mathcal{L}^{YM+gf}(y) \mathcal{O}^{(n)}(x,z)\Big)
&=&
\sum_{\ell=0}^\infty \Big(a\partial_a Z (Z_g Z_A)^\ell\Big) \mathcal{O}^{(n)}_\ell(x,z,A)
\nonumber\\&&\hspace*{-3.5cm}=~
\sum_{\ell=0}^\infty \Big(a\partial_a Z Z^{-1} + \ell a\partial_a\ln(Z_gZ_A) \Big) Z (Z_g Z_A)^\ell \mathcal{O}^{(n)}_\ell(x,z,A)
\nonumber\\&&\hspace*{-3.5cm}=~
\left(a\partial_a Z Z^{-1} +a\partial_a\ln(Z_gZ_A)\int d^dy A(y)\frac{\delta}{\delta
A(y)}\right)[\mathcal{O}^{(n)}(x,z)]\,.
\end{eqnarray}
Adding the second pair counterterm, and taking into account that
\begin{align}
\gamma_A &=(\beta(a)\partial_a-2\gamma_A\xi\partial_\xi) \ln Z_A\,, \qquad\qquad \gamma_g=\beta(a)\partial_a \ln Z_g\,,
\notag\\
\mathbb{H}(a)&=-M\frac{d}{dM} Z Z^{-1}+2\gamma_q=-(\beta(a)\partial_a-2\gamma_A\xi\partial_\xi) Z Z^{-1}+2\gamma_q
\end{align}
one obtains
\begin{align}\label{SRE3}
\VEV{\delta_D S_R\, [\mathcal{O}^{(n)}]\,[\mathcal{O}^{(\bar n)}]}_{\mathcal{L}}
& = -\frac{\beta(a)}{a} \int\!d^d y \VEV{\left[\mathcal{L}^{YM+gf}(y)\, \mathcal{O}^{(n)}\mathcal{O}^{(\bar n)}\right]}
\notag\\
&\hspace*{4mm}
- \left(\mathbb{H}^{(z)}(a)+ \mathbb{H}^{(w)}(a) \right)\VEV{[\mathcal{O}^{(n)}]\,[\mathcal{O}^{(\bar n)}]}
\notag\\
&\hspace*{4mm} + 4 \left(\gamma_q + 4 \gamma_A \xi\partial_\xi \ln Z_q \right)\VEV{[\mathcal{O}^{(n)}]\,[\mathcal{O}^{(\bar n)}]}
\notag\\
&\hspace*{4mm} + \left(\gamma_g + \gamma_A + 2\gamma_A\xi\partial_\xi \ln Z_A\right)
\VEV{\int\!d^dy\, A(y) \frac{\delta}{\delta A(y)}[\mathcal{O}^{(n)}]\,[\mathcal{O}^{(\bar n)}]},
\end{align}
where, to save space, we do not show arguments of the light-ray operators $[\mathcal{O}^{(n)}] = [\mathcal{O}^{(n)}](0,z)$,
$[\mathcal{O}^{(\bar n)}] = [\mathcal{O}^{(\bar n)}](x,w)$.
Note that the expression in the third line exactly cancels the contribution of quark EOM operators, Eq.~\eqref{SRE1},
and the last contribution cancels with the gluon EOM terms in~\eqref{Nt}.
Thus Eq.~\eqref{SRE} is indeed reproduced, as expected, with the identification (cf. Appendix~B)
\begin{align}
a \partial_a\mathcal{G}(x;z,w)
&= \int d^d y \VEV{ \left[\mathcal{L}^{YM+gf}(y) \mathcal{O}^{(n)}(0,z) \mathcal{O}^{(\bar n)}(x,w)\right]}.
\end{align}
We stress that this term does not contribute to the SWI at the critical point.
Also, since all singular terms in $\epsilon$ have to cancel, we could drop them from the beginning and only
consider finite contributions. This cancellation is rather nontrivial on a diagrammatic level. We have demonstrated
how it works for the SWI, but we will simply assume of in the analysis of CWI in the next section.

%%%%%%%%%%%%%%%%%%%%%%%%%%%%%%%%%%%%%%%%%%%%%%%%%%%%%%%%%%%%%%%%%%%%%%%%%%%%%%%%%%%%%%%%%%%%%%%%%%%%%%%%%%%%%%%%%%%%%%
\subsection{Conformal Ward Identity}
%%%%%%%%%%%%%%%%%%%%%%%%%%%%%%%%%%%%%%%%%%%%%%%%%%%%%%%%%%%%%%%%%%%%%%%%%%%%%%%%%%%%%%%%%%%%%%%%%%%%%%%%%%%%%%%%%%%%%%

The two terms on the l.h.s. of the conformal Ward identity (CWI), Eq.~\eqref{SCWI},
correspond to the variation of the light-ray operators. The first one can be expressed in terms of $S_+$,
\begin{align}
\delta_K [\mathcal{O}^{(n)}](0,z)  &= Z\delta_K \mathcal{O}^{(n)}(0,z)= 2(n\bar n)Z S_+^{(\epsilon)}\mathcal{O}^{(n)}(0,z)
=
2(n\bar n) Z S_+^{(\epsilon)}Z^{-1}[\mathcal{O}^{(n)}(0,z)]\,,
\end{align}
where $S_+^{(\epsilon)}=S_+^{(0)}-\epsilon(z_1+z_2)$, the term
$-\epsilon(z_1+z_2)$ is due to the modification of the quark scaling dimension
$\Delta_q = \frac32 -\epsilon$, cf.~\eqref{scalingdimension}.
The product $Z \,S_+^{(\epsilon)}\,Z^{-1}$ can be rewritten after some algebra (see Ref.~\cite{Braun:2013tva}) as
\begin{align}
 Z \,S_+^{(\epsilon)}\,Z^{-1}&= S_+^{(\epsilon)}-\frac12\int_0^{a}\frac{du}{u}\Big[\mathbb{H}(u),z_1+z_2\Big]+ \ldots
\notag\\&= S_+^{(\epsilon)}-\frac12 a\, [\mathbb{H}^{(1)},z_1+z_2] - \frac14 a^2 \,[\mathbb{H}^{(2)},z_1+z_2] + \mathcal{O}(a^3)+\ldots
\end{align}
where the ellipses stand for the singular $1/\epsilon$ terms.
As discussed above, the explicit expression for the singular contributions is not needed since they must cancel
in the final result.

It is easy to show that the conformal variation of the second light-ray operator retains its leading order
form (for our choice $(x\cdot\bar n)=0$)
\begin{align}
\delta_K [\mathcal{O}^{(\bar n)}](x,w)=-x^2(\bar n\cdot \partial_x) [\mathcal{O}^{(\bar n)}](x,w)\,.
\end{align}
Thus the CWI takes the form
\begin{align}\label{CWI1}
\Big( 2(n\bar n)Z S_+^{(\epsilon)} Z^{-1} -x^2(\bar n\cdot \partial_x)\Big)\mathcal{G}(x;z,w)
%\VEV{[\mathcal{O}^{(n)}](0,z)\,[\mathcal{O}^{(\bar n)}](x,w)}
&=\VEV{\delta_K S_R\, [\mathcal{O}^{(n)}](0,z)\,[\mathcal{O}^{(\bar n)}](x,w)}
\notag\\
&= \int d^dy \, 2(\bar n\cdot y)\,\VEV{\mathcal{N}(y)[\mathcal{O}^{(n)}](0,z)\,[\mathcal{O}^{(\bar n)}](x,w)},
\end{align}
where we have discarded the term due to the BRST operator $\partial^\rho \mathcal{B}_\rho$~\eqref{DC} as it does not
contribute to gauge-invariant correlation functions.

The contribution due to the quark EOM reads
\begin{align}
\label{quarkEOM}
\VEV{\delta_K S_R\, [\mathcal{O}^{(n)}]\,[\mathcal{O}^{(\bar n)}]}_{\Omega_q} &=
-2(n\bar n) \gamma_q (z_1+z_2) \mathcal{G}(x;z,w)
%\VEV{[\mathcal{O}^{(n)}](0,z)\,[\mathcal{O}^{(\bar n)}](x,w)}
~+\,\text{singular terms}\,.
\end{align}
Again, the singular terms can be dropped since they must cancel.

The next contribution is due to
\begin{align}\label{NIntegral}
\VEV{\delta_K S_R\, [\mathcal{O}^{(n)}]\,[\mathcal{O}^{(\bar n)}]}_{\mathcal{L}}
& = -\frac{\beta(a)}{a} \int\!d^d y\,2(\bar n\cdot y)
 \VEV{[\mathcal{L}^{YM+gf}(y)]\, [\mathcal{O}^{(n)}]\,[\mathcal{O}^{(\bar n)}]}
\end{align}
Similar to the case of the SWI, the correlation function can be written as the finite part, plus
contributions of pair counterterms corresponding to the contraction of $[\mathcal{L}^{YM+gf}(y)]$ and one of the
light-ray operators. The principal difference is that now we need terms involving the first derivative
of the delta-function in  \eqref{CT}. Such terms cannot be written in terms of the evolution kernel $\mathbb{H}$ and
require a separate calculation.

 It is easy to see that for the case of $\mathcal{O}^{(\bar n)}$ aligned in the
same direction as the parameter in the conformal transformation, $\delta_K=\bar n \cdot \mathbf{K}$,
all pair counterterms vanish as the factor $(\bar n\cdot y)$ under the integral inevitably produces $\bar n^2=0$.
Thus we only need  pair counterterms for the product  $[\mathcal{L}^{YM+gf}(y)][\mathcal{O}^{(n)}](0,z)$
which can be calculated considering the Green function $\vev{\mathcal{O}^{(n)}(0,z)q(p)\bar q(p'))}$
with an insertion of the additional  vertex $\int d^d y\, 2(\bar n\cdot y) \mathcal{L}^{YM+gf}(y)$.
The corresponding contribution to the correlation function of the two light-ray operators is, for a given Feynman
diagram~$D$,
\begin{align}\label{ZACt}
\PCt(D)=-\text{KR}'(D)=  (n\cdot \bar n)Z_D(a) \mathcal{G}_0(x;z,w)\,,
%\vev{\mathcal{O}^{(n)}(0,z)[\mathcal{O}^{(\bar n)}](x,w)}_0,
\end{align}
where $\mathcal{G}_0(x;z,w)$ is the leading-order correlation function \eqref{FInt1}
and the renormalization factor $Z_D(a)$ is an integral operator in $z = \{z_1,z_2\}$ which has the expansion
$$
Z_D(a)=\frac1\epsilon Z_D^{(1)}(a) + \frac1{\epsilon^2} Z_D^{(2)}(a)+\ldots.
$$
Taking into account that $\beta(a)/a = -2\epsilon -2\gamma_g $
we obtain in this way
\begin{align}\label{NIntegral-1}
\VEV{\delta_K S_R\, [\mathcal{O}^{(n)}]\,[\mathcal{O}^{(\bar n)}]}_{\mathcal{L}}
& = -\frac{\beta(a)}{a} \times\,\text{finite~terms} -
2(n\cdot\bar n) \delta S_+(a)\,\mathcal{G}_0(x;z,w)
\notag\\&\hspace*{4mm}+ \text{singular terms} +\ldots
\end{align}
where the operator $\delta S_+(a) = a\, \delta S^{(1)}_+ + a^2 \delta S^{(2)}_++\ldots $ is given by the sum of the simple residues
\begin{align}
\delta S_+(a) = \sum_{D}  Z^{(1)}_D(a)\,.
\end{align}
One should expect that in the final answer $\mathcal{G}_0(x;z,w)$ will be substituted by the complete correlation function
$$
 \mathcal{G}(x;z,w) = \mathcal{G}_0(x;z,w) + \mathcal{G}_1 (x;z,w) +\ldots
$$
so that the same integral operator $\Delta S_+(a)$ appears for the correlation function at any order of
perturbation theory. This property does not hold for the  pair counterterm contributions alone
where, in general, \emph{different} operators $\delta S_+$, $\delta S'_+$ can appear,
\begin{align}
\delta S_+\,\mathcal{G}_0(x;z,w) +
\delta S'_+\, \mathcal{G}_1(x;z,w) +\ldots\,,
\end{align}
etc., and it has to be restored by adding the gluon EOM contributions $\propto (\gamma_A+\gamma_g)\Omega_A$.%
\footnote{Note that $\gamma_A + \gamma_g = 0$ in the background field gauge in which case the product $gA$ is not renormalized.
In this gauge universality of the $\Delta S_+(a)$ operator should hold for pair counterterm contributions alone.}

Summing all contributions and taking into account that
the first contribution in~\eqref{NIntegral} vanishes at the critical point and all singular terms $1/\epsilon$ must cancel,
the CWI~\eqref{SCWI} takes the expected form, \eqref{EQ},
where the operator of special conformal transformation $S_+$ is given by the following expression:
\begin{align}\label{S+}
S_+(a_*) & \equiv S_+^{(0)}+ \Delta S_+
\notag\\
& =S_+^{(\epsilon)}-\frac12\int_0^{a_*}\frac{du}{u}\big[\mathbb{H}(u),z_1+z_2\big]+ \gamma^\ast_q (z_1+z_2) +\delta S_+(a_*)
\notag\\
&=
S_+^{(0)}+\big(\gamma^\ast_q-\epsilon\big) (z_1+z_2)+\delta S_+(a_*)-\frac12\sum_{k=1}^\infty \frac1k a_*^k\big[\mathbb{H}^{(k)},z_1+z_2\big]\,.
\end{align}
Here $\gamma_q^\ast = \gamma_q(a_\ast)$
and it is understood that the shift in the space-time dimension $\epsilon = (4-d)/2$ is written
as an expansion in terms of the critical coupling, $\epsilon=\epsilon(a^*)$.
The role of the term $\big(\gamma^\ast_q-\epsilon\big) (z_1+z_2)$ is to shift the conformal spin of the quark field
to its correct value at the critical point,
\begin{align}\label{add1}
S^{(0)}_++\big(\gamma^\ast_q-\epsilon\big) (z_1+z_2) &=z_1^2\partial_{z_1}+z_2^2\partial_{z_2}+2 j_q^\ast(z_1+z_2)\,,
\notag\\
 2 j_q^\ast &= \Delta_q +\gamma^\ast_q  + \frac12 = 2-\epsilon + \gamma^\ast_q\,.
\end{align}
Taking into account that quantum corrections to $S_0$ are given entirely
in terms of the evolution kernel $\mathbb{H}$, cf.~\eqref{exactS},
it follows from the commutation relation $[S_+,S_-]=2 S_0$ that $[\Delta S_+(a_*), S_{-}] = \mathbb{H}(a_*)$
where $S_- = S_-^{(0)} = -\partial_{z_1}-\partial_{z_2} $ is the generator of translations along the light cone.
This suggests that the correction term $\delta S_+(a_*)$ can be written in the form
\begin{align}\label{Delta+def}
\delta S_+(a_*)=\frac12\big[\mathbb{H}(a_*)-2\gamma_q^\ast\big] (z_1+z_2) + z_{12} \Delta_+(a_*)\,,
\end{align}
where the operator $\Delta_+$ commutes with $S_-$ and anticommutes with the permutation operator of
quark coordinates $\mathbb{P}_{12} f(z_1,z_2)=f(z_2,z_1)$,
\begin{align}
    \mathbb{P}_{12} \Delta_+=-\Delta_+ \mathbb{P}_{12}.
\end{align}
The role of the term $-\gamma^\ast_q(z_1+z_2)$ is to cancel the corresponding term in \eqref{S+}, \eqref{add1} such that
the (gauge-dependent) quark anomalous dimension falls out of the final answer.
We will see that the structure~\eqref{Delta+def} indeed arises naturally in the calculation.

Finally, replacing $\epsilon \mapsto -\gamma_g(a_*) = -\beta_0 a_\ast -\beta_1 a_\ast^2 -\ldots$ we obtain the following expression
for the $\ell$-loop correction to the generator of special conformal transformations:
\begin{align}
 \Delta S_+^{(\ell)} &=
\left(\beta_{\ell-1} + \frac12 \mathbb{H}^{(\ell)}\right) (z_1+z_2) -\frac1{2\ell}\big[\mathbb{H}^{(\ell)},z_1+z_2\big]
 + z_{12} \Delta^{(\ell)}_+\,.
\label{S+loop}
\end{align}

%%%%%%%%%%%%%%%%%%%%%%%%%%%%%%%%%%%%%%%%%%%%%%%%%%%%%%%%%%%%%%%%%%%%%%%%%%%%%%%%%%%%%%%%%%%%%%%%%%%%%%%%%%%%%%%%%%%%%%
\section{Technical details}
%\section{Generator of special conformal transformations $\Delta S_+$}
%%%%%%%%%%%%%%%%%%%%%%%%%%%%%%%%%%%%%%%%%%%%%%%%%%%%%%%%%%%%%%%%%%%%%%%%%%%%%%%%%%%%%%%%%%%%%%%%%%%%%%%%%%%%%%%%%%%%%%

In this section we present technical details of the calculation.
The problem reduces to the calculation of singular contributions to the 1PI Feynman diagrams for the Green function
$$
  \int\! d^dx_1 \!\int\! d^dx_2 \,e^{ip_1x_1-ip_2x_2}  \left\langle q(x_1) \bar q(x_2) \! \int d^d y\,2 (\bar n\cdot y) \mathcal{L}^{YM+gf}(y) \mathcal{O}^{(n)}(0,z)\right\rangle\,.
$$
Note that the counterterms corresponding to the renormalization of the light-ray operator and the Lagrangian insertion
(that we do not need) are disposed of by the usual R-operation.

It is convenient to rewrite
\begin{flalign}\label{LL}
2\!\int d^d y\, (\bar n\cdot y) \mathcal{L}^{YM+gf}(y)&
=2\!\int d^d x(\bar n\cdot  y)\Big[-\frac12A^a_\mu K^{\mu\nu} A^a_\nu
+\mathcal{L}^{YM}_{\text{int}}(y) +  \frac{\xi+1}{2\xi}\partial^\mu\Big( A_\mu (\partial A)\Big)\Big].
\end{flalign}
Here $\mathcal{L}^{YM}_{\text{int}}(y)$ contains the three- and four-gluon interaction vertices  and
\begin{align}
K^{\mu\nu}=g^{\mu\nu} \partial^2  - \partial^\mu\partial^\nu\left(1-\frac1\xi\right)\,.
\end{align}
The last term in \eqref{LL} can be represented in the  form $A_\mu (\partial A)=\xi(\bar c D_\mu
c-\mathcal{B}_\mu)$.  The operator $\mathcal{B}_\mu$ is a BRST variation  and, hence,  does not contribute to the correlation function.
Omitting this term one gets for~\eqref{LL}
\begin{align}\label{LM}
\int d^d y (\bar n\cdot y)\Big[- A^a_\mu K^{\mu\nu} A^a_\nu
+2\mathcal{L}^{YM}_{\text{int}}(y) +  \left(1+\xi\right)\partial^\mu\big(\bar c D_\mu c\big)\Big]\,.
\end{align}
This insertion generates  two-, three- and four-gluon vertices as well as ghost-antighost and ghost-antighost-gluon
vertices.
An insertion of the two-gluon effective vertex in the gluon line results in the following effective
propagator
\begin{align}\label{eff-prop}
\begin{minipage}{3.2cm}{\includegraphics[width=3.2cm]{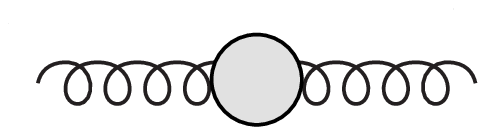}}\\ \end{minipage}
\begin{minipage}{0.5cm}{$=$}\end{minipage}
\begin{minipage}{5.5cm}{\includegraphics[width=6cm]{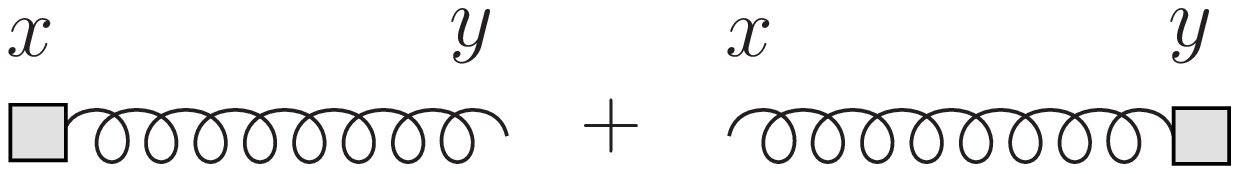}\\}\end{minipage}
\end{align}
where the gray boxes at the endpoints stand for the multiplication by the corresponding coordinates, $(\bar n\cdot x)$ and $(\bar n \cdot y)$, respectively.
Such insertions violate translation invariance and the main trick is to move them either to the external quark lines or to the quark positions in the light-ray
operator in which case the corresponding singular contributions can be related to the evolution kernel. Examples will be given below.
A shift $x\mapsto y $ corresponds to the simple rewriting
\begin{align}
   \bar n\cdot x = \bar n\cdot(x-y) + \bar n\cdot y
\end{align}
that can be represented diagrammatically as
\begin{align}\label{shift}
\begin{minipage}{9cm}{\includegraphics[width=9cm]{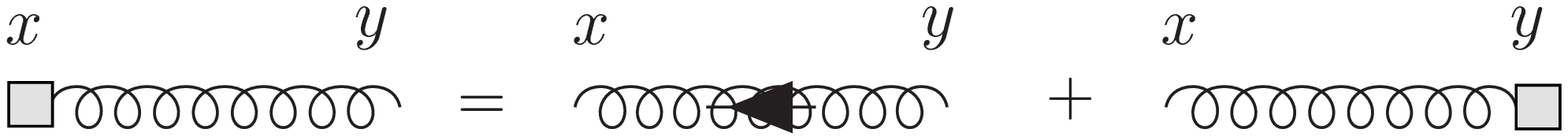}}\end{minipage}
\end{align}
where the gluon propagator with a thick arrow (in Feynman gauge) is defined as%
\footnote{All expressions are given in Euclidean space}
\begin{align}\label{gluonarrow}
\begin{minipage}{3.0cm}{\includegraphics[width=2.9cm]{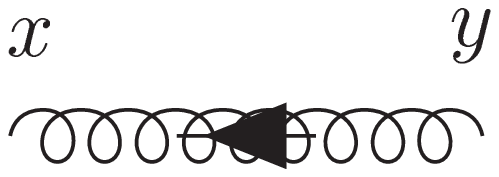}}\\ \end{minipage}
\begin{minipage}{0.5cm}{$=$}\end{minipage}
\bar n\cdot (x-y)\wick{1}{<1A_\mu(x)>1A_\nu(y)}
%\mathcal{D}_{\mu\nu}(x-y)
~=~2 ig_{\mu\nu}\int\frac{d^d k}{(2\pi)^d} e^{-ik\cdot(x-y)}\frac{(\bar n\cdot k)}{k^4}\,.
\end{align}
It is sometimes convenient to move the coordinate insertion along the quark lines and/or along the gauge link so we
also introduce notations
\begin{align}\label{quarkarrow}
&\begin{minipage}{3.0cm}{\includegraphics[width=2.9cm]{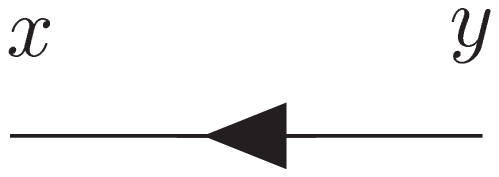}}\\ \end{minipage}
\begin{minipage}{0.5cm}{$=$}\end{minipage}
\bar n\cdot (x-y) \wick{1}{<1q(x)>1{\overline{q}(y)}}
%\mathcal{S}(x-y)
~=~ -\,\int\frac{d^d k}{(2\pi)^d} e^{-ik\cdot(x-y)}\frac{\slashed{k}\slashed{\bar n}\slashed{ k}}{k^4}\,,
\notag\\
&\begin{minipage}{3.0cm}{\includegraphics[width=2.9cm]{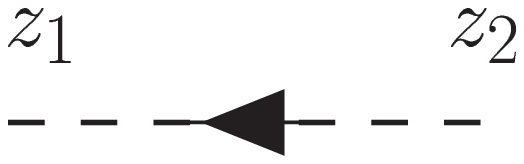}}\\ \end{minipage}
\begin{minipage}{0.5cm}{$=$}\end{minipage}
(\bar n\cdot n) z_{12}\, [z_1n,z_2n]\,,
\end{align}
where
\begin{align}
[z_1n,z_2n]=\text{Pexp} \left\{ig z_{12}\int_0^1 du\, n^\mu A_\mu(z_{21}^u n)\right\}\,.
\end{align}
%

%%%%%%%%%%%%%%%%%%%%%%%%%%%%%%%%%%%%%%%%%%%%%%%%%%%%%%%%%%%%%%%%%%%%%%%%%%%%%%%%%%%%%%%%%%%%%%%%%%%%%%%%%%%%%%%%%%%%%%
\subsection{One-loop calculation}
%%%%%%%%%%%%%%%%%%%%%%%%%%%%%%%%%%%%%%%%%%%%%%%%%%%%%%%%%%%%%%%%%%%%%%%%%%%%%%%%%%%%%%%%%%%%%%%%%%%%%%%%%%%%%%%%%%%%%%

\begin{figure}[t]
\centerline{\includegraphics[width=0.45\linewidth]{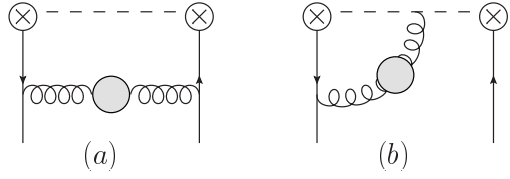}}
\caption{\sf One-loop Feynman diagrams for the quantum correction to the generator of special conformal
transformations.}
\label{fig:oneloop}
\end{figure}

Only the two-gluon effective vertex, $-\frac12(\bar n \cdot x)A^\mu K_{\mu\nu}A^\nu$, is relevant to this accuracy.
There are two one-loop diagrams, shown in Fig.~\ref{fig:oneloop}, and the diagram symmetric to the one in Fig.~\ref{fig:oneloop}b
with the gluon attached to the quark field. The addition of such symmetric contributions is always implied.

%%%%%%%%%%%%%%%%%%%%%%%%%%%%%%%%%%%%%%%%%%%%%%%%%%%%%%%%%%%%%%%%%%%%%%%%%%%%%%%%%%%%%%%%%%%%%%%%%%%%%%%%%%%%%%%%%%%%%%
%%%%%%%%%%%%%%%%%%%%%%%%%%%%%%%%%%%%%%%%%%%%%%%%%%%%%%%%%%%%%%%%%%%%%%%%%%%%%%%%%%%%%%%%%%%%%%%%%%%%%%%%%%%%%%%%%%%%%%

The first diagram, Fig.~\ref{fig:oneloop}a, corresponds to the attachment of the $(\bar n \cdot x_1)$ or $(\bar n \cdot x_2)$
factor to the external (anti)quark line,
\begin{align*}
\begin{minipage}{11.2cm}
{\includegraphics[width=11.2cm]{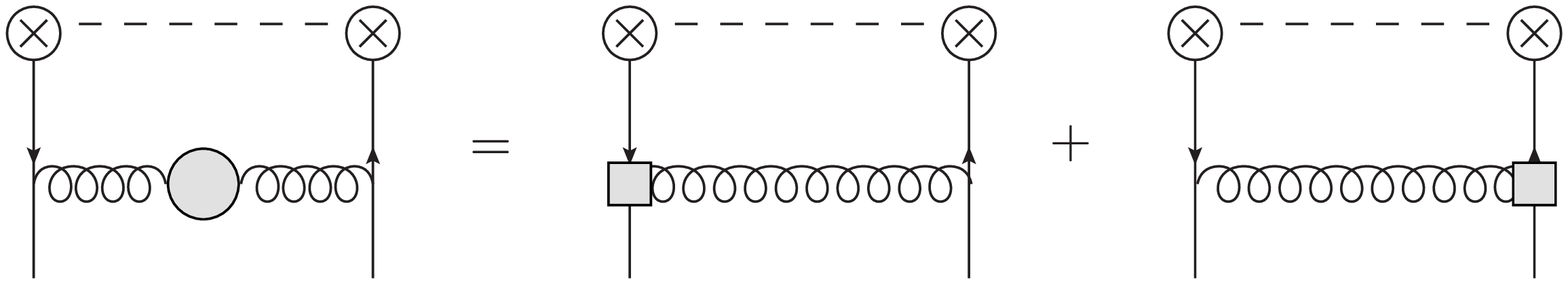}}
\end{minipage}
\end{align*}
The  contribution of such a diagram to $\delta S^{(1)}_+$, Eq.~\eqref{ZACt},
 is related to the contribution of the corresponding diagram
(without $(\bar n\cdot x_k)$ insertion) to the evolution kernel, $\mathbb{H}^{(1)}_{(a)}$. For the left and the right diagrams shown above, one gets
\begin{align}
\delta S^{(1)}_{+(a)L}= \frac12 \mathbb{H}^{(1)}_{(a)} z_1, && \delta S^{(1)}_{+(a)R}= \frac12 \mathbb{H}^{(1)}_{(a)} z_2\,,
\label{Z1a}
\end{align}
respectively. The factor 1/2 is due to the definition of $\mathbb{H}$ that involves a derivative $a\partial_a$ of the corresponding
$Z$-factor, see Eq.~\eqref{H-Z}.  For a generic $\ell$-loop diagram $D$ of this type one gets for the sum of contributions with
$(\bar n\cdot x_k)$ attachments to the quark and the antiquark line
\begin{align}\label{ZHrelation}
\delta S_{+D}^{(1,\ell)} = \frac1{2\ell}\mathbb{H}^{(\ell)}_D (z_1+z_2)\,.
\end{align}
Thus such diagrams do not require a separate calculation.

The second diagram,  Fig.~\ref{fig:oneloop}b, involves integration over the position of the gluon emitted from the gauge link on the light-cone
\begin{align}
  \int_0^1 du\, n^\mu g A_\mu(z_{21}^u n)\,.
\label{gaugelink1}
\end{align}
It can be represented as a sum of three contributions:
\begin{align*}
\begin{minipage}{14.2cm}
{\includegraphics[width=14.2cm]{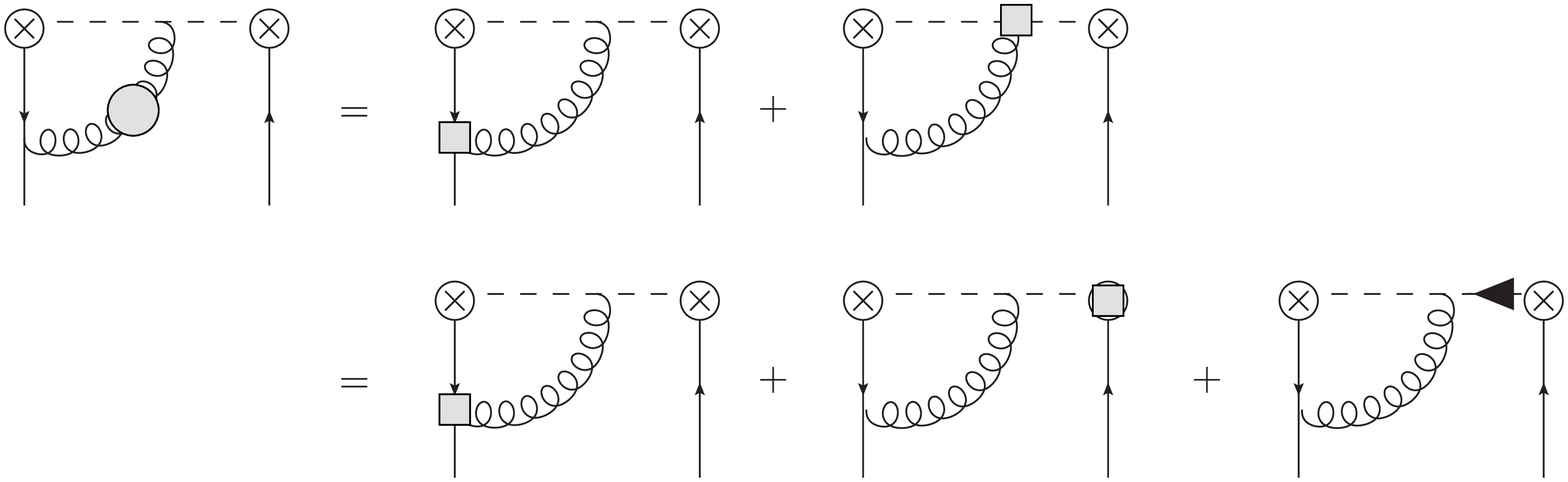}}\\
\end{minipage}
\end{align*}
The first and the second term are of the same type as above and sum up to
\begin{align}
\delta S^{(1)}_{+(b1+b2)}= \frac12 \mathbb{H}^{(1)}_{(b)} (z_1+z_2)\,,
\end{align}
where $\mathbb{H}^{(1)}_{(b)}$ stands for the corresponding contribution to the evolution kernel.
Taking into account the symmetric contribution with the gluon attached to another  quark line,
and adding the contribution of the diagram in Fig.~\ref{fig:oneloop}a, Eq.~\eqref{Z1a}, we obtain for the sum of these terms
\begin{align}
                   \frac12 \big(\mathbb{H}^{(1)}-2\gamma_q^{(1)}\big) (z_1+z_2)\,,
\end{align}
with the complete one-loop evolution kernel \eqref{Honeloop}, which is exactly the anticipated first term in Eq.~\eqref{Delta+def}.

The third term involves the insertion of the gluon and quark coordinates in the gauge link
\begin{align}
 \bar n \cdot(z_{21}^u n -z_2 n) = u z_{12} (n\bar n)\,.
\end{align}
The same diagram without this insertion (i.e. without the thick arrow) gives rise to the contribution to the
evolution kernel~\cite{Balitsky:1987bk}:
\begin{align}
\mathbb{H}_{(b)}^{(1)}f(z_1,z_2)&= \int_0^1d\alpha\, h_{(b)}^{(1)}(\alpha) \Big[f(z_1,z_2)-f(z_{12}^\alpha,z_2)\Big]\,,
\qquad h_{(b)}^{(1)}(\alpha) = 4C_F \frac{\bar\alpha}{\alpha}\,.
\end{align}
The characteristic structure $\sim[f(z_1,z_2)-f(z_{12}^\alpha,z_2)]$ corresponding to the ``plus'' distribution in momentum space
can be traced to the integration over the gluon position on the light-cone such that the above answer arises from the
representation
\begin{align}
\mathbb{H}_{(b)}^{(1)}f(z_1,z_2)& = \int_0^1d\alpha\, h_{(b)}^{(1)}(\alpha) \int_0^1\!du \frac{d}{du} f(z_{12}^{\alpha\bar u},z_2)\,,
\end{align}
where $u$ is the gauge link variable as in \eqref{gaugelink1}, as an intermediate step.
The diagram ``with an arrow'' is given, therefore, by the same expression with an insertion of $u z_{12} (n\bar n)$
and adding the factor 1/2 due to a different normalization
%\footnote{Extra factor 1/2 is due to a different normalization.}
%
\begin{align}
\delta S^{(1)}_{+(b3)}f(z_1,z_2) &= z_{12} (n\bar n) \frac12\int_0^1d\alpha\, h_{(b)}^{(1)}(\alpha) \int_0^1\!du u \frac{d}{du} f(z_{12}^{\alpha\bar u},z_2)
\notag\\&=
 z_{12} (n\bar n) \frac12 \int_0^1\!du \int_0^1d\alpha\, h_{(b)}^{(1)}(\alpha) \Big[f(z_1,z_2)-f(z_{12}^{\alpha\bar u},z_2)\Big]
\notag\\&=
 z_{12} (n\bar n) \frac12 \int_0^1\!d\alpha \left(\int_\alpha^1 \frac{du}{u}\, h_{(b)}^{(1)}(u)\right) \Big[f(z_1,z_2)-f(z_{12}^{\alpha},z_2)\Big].
\end{align}
The symmetric diagram with the gluon attached to the quark instead of the antiquark gives the same contribution up to
a replacement $z_1\leftrightarrow z_2$ so that in the prefactor $z_{12} \to -z_{12}$. The symmetric contribution is, therefore,
effectively subtracted and one obtains in the sum
\begin{align}
 \delta S^{(1)}_{+(b3+{\rm sym})}f(z_1,z_2) &=  z_{12} (n\bar n) \frac12 \int_0^1\!d\alpha \left(\int_\alpha^1 \frac{du}{u}\, h_{(b)}^{(1)}(u)\right) \Big[f(z_1,z_{21}^\alpha)-f(z_{12}^{\alpha},z_2)\Big]
\notag\\&=
 z_{12} (n\bar n) 2C_F \int_0^1\!d\alpha \left(\frac{\bar\alpha}{\alpha} + \ln\alpha\right) \Big[f(z_1,z_{21}^\alpha)-f(z_{12}^{\alpha},z_2)\Big]\,,
\end{align}
reproducing the result quoted in \eqref{Delta+1}~\cite{Braun:2014vba}.

It is easy to show that this kind of relation between the diagrams with and without the arrow on the gauge link is general
and true in all orders, the reason being that integration over the position of the gluon field in the light-ray operator
does not interfere with the separation of singular parts. For a generic $\ell$-loop contribution of this type one obtains
\begin{align*}
\begin{minipage}{3.0cm}
{\includegraphics[width=3.0cm]{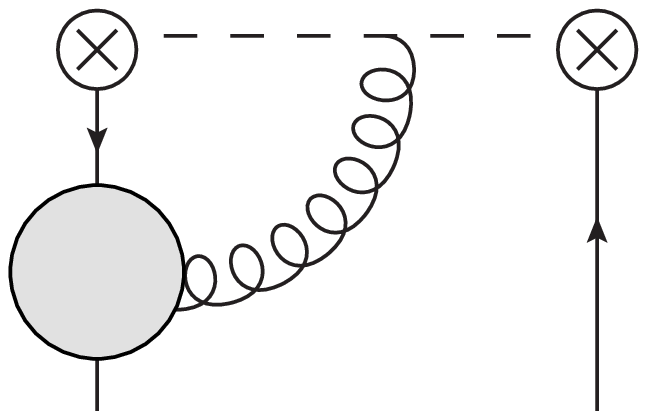}}\\
\end{minipage}
& :\quad\mathbb{H}_{(D)}^{(\ell)}f(z_1,z_2)=\int_0^1d\alpha\, h_{(D)}^{(\ell)}(\alpha) \Big[f(z_1,z_2)-f(z_{12}^\alpha,z_2)\Big],
\notag\\
\begin{minipage}{3.0cm}
{\includegraphics[width=3.0cm]{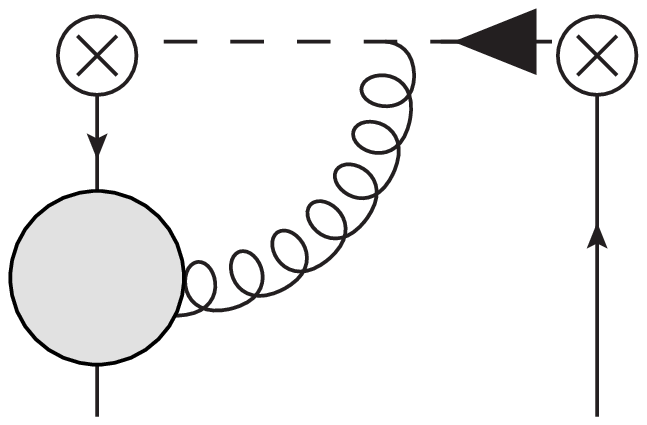}}\\
\end{minipage}
& :\quad
\delta S^{(1)}_{+(D)}f(z_1,z_2)=\frac{z_{12}}{2\ell} (n\bar n) \int_0^1\!du\! \int_0^1\!d\alpha\,
h_{(D)}^{(\ell)}(\alpha) \Big[f(z_1,z_2)-f(z_{12}^{\alpha\bar u},z_2)\Big],
\end{align*}
with the same function $h_{(D)}^{(\ell)}(\alpha)$. Thus diagrams with an arrow on the gauge link do not require a separate calculation
as well.

%%%%%%%%%%%%%%%%%%%%%%%%%%%%%%%%%%%%%%%%%%%%%%%%%%%%%%%%%%%%%%%%%%%%%%%%%%%%%%%%%%%%%%%%%%%%%%%%%%%%%%%%%%%%%%%%%%%%%%
\begin{figure}[t]
\centerline{\includegraphics[width=0.99\linewidth]{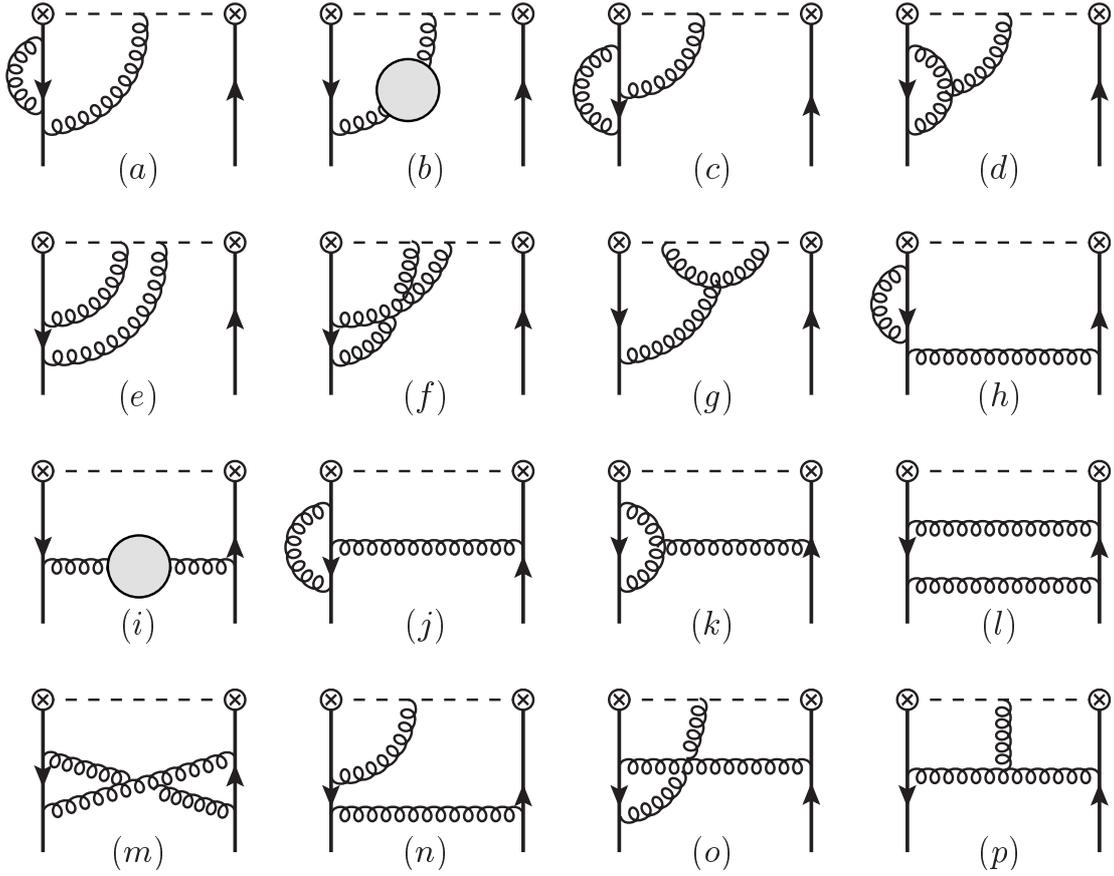}}
\caption{\sf Feynman diagrams of different topologies contributing to the two-loop
evolution equation and the two-loop deformation of the generator of special conformal
transformations for flavor-nonsiglet leading twist operators.\label{fig:Diagrams}}
\end{figure}
%%%%%%%%%%%%%%%%%%%%%%%%%%%%%%%%%%%%%%%%%%%%%%%%%%%%%%%%%%%%%%%%%%%%%%%%%%%%%%%%%%%%%%%%%%%%%%%%%%%%%%%%%%%%%%%%%%%%%%

%%%%%%%%%%%%%%%%%%%%%%%%%%%%%%%%%%%%%%%%%%%%%%%%%%%%%%%%%%%%%%%%%%%%%%%%%%%%%%%%%%%%%%%%%%%%%%%%%%%%%%%%%%%%%%%%%%%%%%
\subsection{Two loop calculation}
%%%%%%%%%%%%%%%%%%%%%%%%%%%%%%%%%%%%%%%%%%%%%%%%%%%%%%%%%%%%%%%%%%%%%%%%%%%%%%%%%%%%%%%%%%%%%%%%%%%%%%%%%%%%%%%%%%%%%%

To two-loop accuracy we have to have to take into account the Feynman diagrams of 16 different topologies shown in
Fig.~\ref{fig:Diagrams}.  Each of these diagrams gives rise to several contributions to $\Delta S_+$ corresponding
to a replacement of either one of the gluon propagators by the effective propagator~\eqref{eff-prop}, or of the
three-gluon (and quark-antiquark gluon) vertex by the corresponding effective vertex~\eqref{LM}.
Below we tacitly imply using Feynman gauge, $\xi=1$.

The ``QED type'' diagrams in Fig.~\ref{fig:Diagrams} (a-c), (e),(f), (h-j), (l-o) can be calculated using the same
strategy as the one-loop diagrams considered above.  Let us
illustrate this procedure on the example of the diagram Fig.~\ref{fig:Diagrams}(j). Inserting the effective gluon
propagators one obtains four contributions:
\begin{align*}
%\label{S1L}
\begin{minipage}{15.0cm}
{\includegraphics[width=15.0cm]{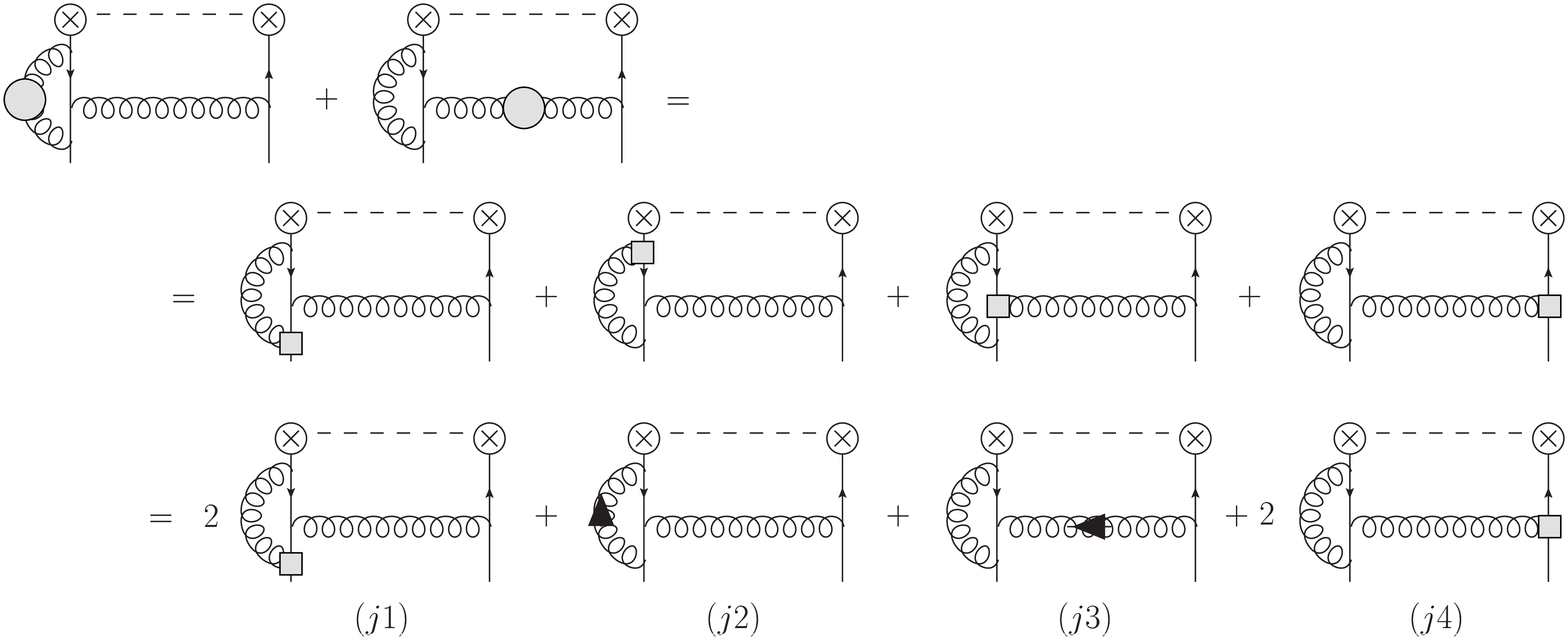}}\\
\end{minipage}
\end{align*}
The first and the last one, (j1) and (j4), combine to
\begin{align}
   \Big(\delta S_{+}^{(2)}\Big)_{j1+j4} = \frac12 \mathbb{H}_{(j)}^{(2)}(z_1+z_2)\,,
\end{align}
where $\mathbb{H}_{(j)}^{(2)}$ is the contribution of the diagram in  Fig.~\ref{fig:Diagrams}(j) to the evolution kernel.
The remaining diagrams (j2) and (j3) with a modified gluon propagator \eqref{gluonarrow} have to be calculated explicitly. The result
can be found in Appendix~\ref{App:diagrams}.

\begin{figure}[h]
\includegraphics[width=13.7cm]{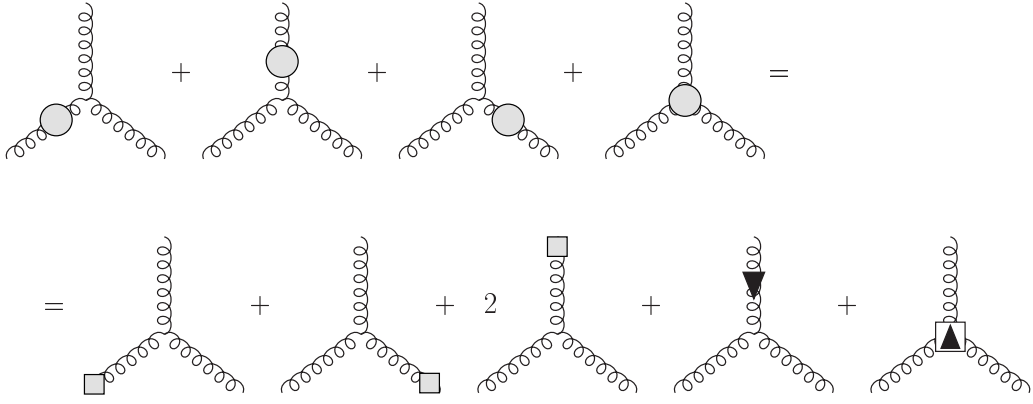}
\caption{\sf Rearrangement of three-gluon vertex insertions combined with effective propagators.  \label{fig:3GVertex}}
\end{figure}

The diagrams which contain the three-gluon vertex can be handled in the following way.
It is convenient to consider the sum of contributions with the effective vertex and effective gluon propagators and rearrange it as
shown in Fig.~\ref{fig:3GVertex}.
%
%%%%%%%%%%%%%%%%%%%%%%%%%%%%%%%%%%%%%%%%%%%%%%%%%%%%%%%%%%%%%%%%%%%%%%%%%%%%%%%%
%
%
%%%%%%%%%%%%%%%%%%%%%%%%%%%%%%%%%%%%%%%%%%%%%%%%%%%%%%%%%%%%%%%%%%%%%%%%%%%%%%%%%%
%
Here the gray blobs in the diagrams in the first row stand for insertions generated by \eqref{LM}. Their sum
can be rewritten as shown in the second row where the white box with an arrow denotes a new vertex
\begin{align}
%\label{S1L}
\begin{minipage}{3.3cm}
{\includegraphics[width=3.0cm]{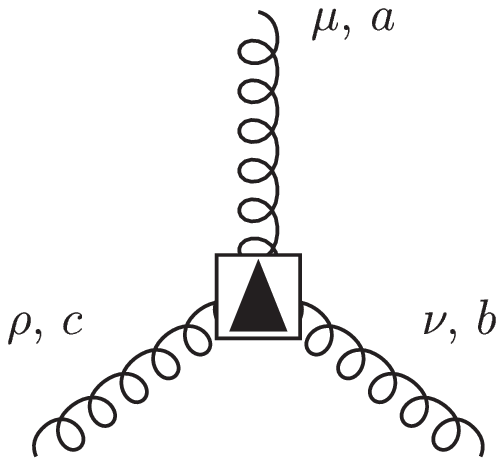}}
\end{minipage}&
%\Delta V(\mu,a; \nu,b; \rho,c)=
g f^{abc}\Big(g^{\mu\rho}\bar n^\nu - g^{\mu\nu}\bar n^\rho\Big).
\end{align}
Note that this vertex is symmetric under the interchange of the lower pair of gluons, $(\nu,b)\leftrightarrow (\rho,c)$,
but the line $(\mu,a) $ is distinguished, hence an arrow in the notation. This special direction has to be chosen in such a way
that the contributions with the insertion of  $(\bar n\cdot x_k)$ factors (gray boxes) in the external lines combine to produce
a term $\sim \mathbb{H}_{(D)}^{(2)}(z_1+z_2)$.  For example, the contribution of the diagram in Fig.~\ref{fig:Diagrams}(k) can be
split in the following five terms:
\begin{align*}
\begin{minipage}{15.0cm}
{\includegraphics[width=15.0cm]{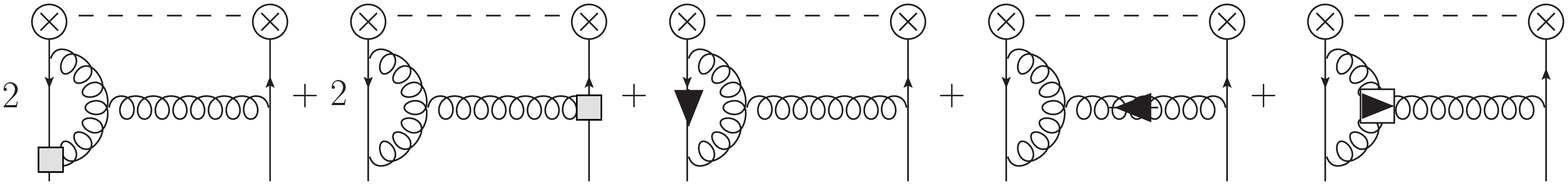}}\\
\end{minipage}
\end{align*}
The first two contributions give rise to the term  $\frac12 \mathbb{H}_{(k)}^{(2)}(z_1+z_2)$ where $\mathbb{H}_{(k)}^{(2)}$ is the
contribution of the diagram in Fig.~\ref{fig:Diagrams}(k) to the evolution kernel, and the remaining three have to be
calculated explicitly, see Appendix~\ref{App:diagrams}.

Finally there are four diagrams with self-energy insertions, Fig.~\ref{fig:Diagrams}(a),(b),(h),(i). It is easy to see that
\begin{align*}
\begin{minipage}{8.0cm}
{\includegraphics[width=8.0cm]{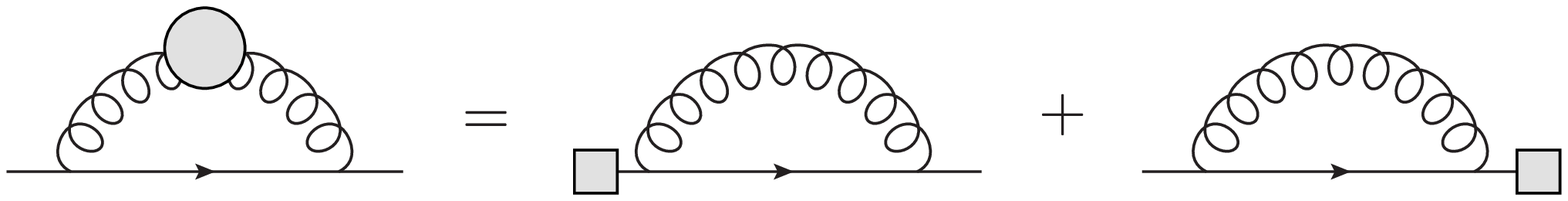}}\\
\end{minipage}
\end{align*}
and also
\begin{align*}
\begin{minipage}{11.0cm}
{\includegraphics[width=11.0cm]{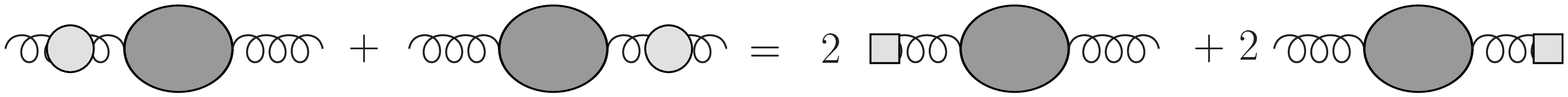}}\\
\end{minipage}
\end{align*}
where the dark oval corresponds to the sum of the contributions of quark, gluon and ghost loops.
Using these replacement rules  one obtains immediately,  e.g., for the diagrams in Fig.~\ref{fig:Diagrams}(h) and Fig.~\ref{fig:Diagrams}(i),
\begin{align}
 \Big(\delta S_{+}^{(2)}\Big)_{(h)} &=\frac12\mathbb{H}^{(2)}_{(h)}(z_1+z_2)+\frac14 \Big(z_1 \mathbb{H}^{(2)}_{(h)}-\mathbb{H}^{(2)}_{(h)} z_2\Big)\,,
\notag\\
 \Big(\delta S_{+}^{(2)}\Big)_{(i)} &=\frac12\mathbb{H}^{(2)}_{(i)}(z_1+z_2)\,,
\end{align}
respectively, where $\mathbb{H}^{(2)}_{(h,i)}$ are the corresponding contributions to the evolution kernel.

Finally, one has to consider insertions of \eqref{LM} in the self-energy blob itself. For the gluon loop one obtains
\begin{align}
%\label{S1L}
\begin{minipage}{13.0cm}
{\includegraphics[width=12.5cm]{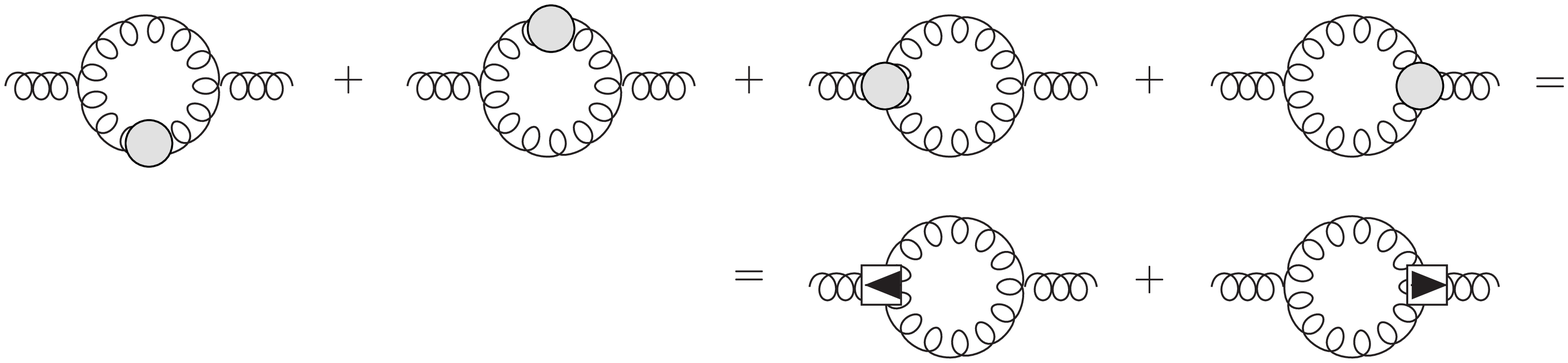}}
\end{minipage}
\end{align}
It can be checked, however, that this contribution is cancelled identically by the similar
diagrams with the ghost-gluon vertex insertions and the insertions in ghost propagators
so that the insertions inside self-energy diagrams can be omitted altogether.

The rest of the calculation is relatively straightforward.
The complete results for the contribution of each Feynman diagram in  Fig.~\ref{fig:Diagrams} to the conformal anomaly and
the evolution kernel are presented in  Appendix~\ref{App:diagrams}.

%%%%%%%%%%%%%%%%%%%%%%%%%%%%%%%%%%%%%%%%%%%%%%%%%%%%%%%%%%%%%%%%%%%%%%%%%%%%%%%%%%%%%%%%%%%%%%%%%%%%%%%%%%%%%%%%%%%%%%
\section{Final results}
%%%%%%%%%%%%%%%%%%%%%%%%%%%%%%%%%%%%%%%%%%%%%%%%%%%%%%%%%%%%%%%%%%%%%%%%%%%%%%%%%%%%%%%%%%%%%%%%%%%%%%%%%%%%%%%%%%%%%%

Let us start with the evolution kernel
\begin{align}
\mathbb{H}(a) &= a\, \mathbb{H}^{(1)}+ a^2\, \mathbb{H}^{(2)}+\ldots.
\end{align}
The one-loop result reads~\cite{Balitsky:1987bk}
\begin{align}
\mathbb{H}^{(1)} f(z_1,z_2) &= 4C_F\biggl\{\frac12 f(z_1,z_2)+
\int_0^1d\alpha\frac{\bar\alpha}{\alpha}\Big[2f(z_1,z_2)-f(z_{12}^\alpha,z_2)-f(z_1,z_{21}^\alpha)\Big]
\notag\\
&\quad -\int_0^1d\alpha\int_0^{\bar\alpha} d\beta f(z_{12}^\alpha,z_{21}^\beta)
\biggr\},
\end{align}
where $f(z_1,z_2)$ is a test function,
and the two-loop kernel~\cite{Braun:2014vba} can be written in the form
\begin{align}
\mathbb{H}^{(2)} f(z_1,z_2) &= 4\biggl\{ X f(z_1,z_2)+
\int_0^1d\alpha\frac{\bar\alpha}{\alpha}h(\alpha) \Big[2f(z_1,z_2)-f(z_{12}^\alpha,z_2)-f(z_1,z_{21}^\alpha)\Big]
\notag\\
&\quad
+
\int_0^1d\alpha\int_0^{\bar\alpha} d\beta \Big[\chi(\alpha,\beta)+
 \chi^{\mathbb{P}}(\alpha,\beta) \mathbb{P}_{12}\Big]f(z_{12}^\alpha,z_{21}^\beta)\biggr\}\,.
\end{align}
Here $\mathbb{P}_{12} f(z_1,z_2)=f(z_2,z_1)$ is the permutation operator and%
\footnote{The factor 1/2 in the second line of the expression for $\chi(\alpha,\beta)$ (shown in red) was missed in Ref.~\cite{Braun:2014vba}.}
\begin{align}
X & =\frac{13}{12}C_F\beta_0 + C_FC_A\left(6\zeta(3)-\frac23\pi^2+\frac{13}6\right)+2C^2_F\left(-6\zeta(3)+\frac13\pi^2+\frac{21}8\right),
\notag\\
h(\alpha)&= C_F\beta_0 \left(\ln\bar\alpha+\frac53\right)
         - C_FC_A\frac13\left({\pi^2}-4\right)
          - 2C^2_F \ln\bar\alpha\left(\frac32-\ln\bar\alpha +\frac{1+\bar\alpha}{\bar\alpha}\ln\alpha\right) ,
\notag\\
\chi(\alpha,\beta) &=-C_F\beta_0 \left[\ln(1-\alpha-\beta)+\frac{11}3\right]
-
2 C_FC_A \left[\Li_2(\tau)-\Li_2(1)+\frac12\ln^2\bar\tau -\frac1\tau\ln\bar\tau+\frac53\right]
\notag\\
&\quad +2 C^2_F\left[2\Li_2(\tau)+\ln^2\bar\tau + {\color{red}\frac12}\ln\tau-\frac{1+\bar\tau}{\tau}\ln\bar\tau+\varphi(\alpha,\beta)\right],
\notag\\
\chi^{\mathbb{P}}(\alpha,\beta)  & =2C_F\left(C_F-\frac12 C_A\right)\Big(\ln^2\bar\tau-2\tau\ln\bar\tau\Big),
\end{align}
where $\tau=\alpha\beta/\bar\alpha\bar\beta$.
%\footnote{The factor 1/2 in the second line of the expression for $\chi(\alpha,\beta)$ (shown in red) was missed in Ref.~\cite{Braun:2014vba}.}
The function $\varphi(\alpha,\beta)$ is defined as
\begin{align}
\varphi(\alpha,\beta)&=-\Big[\frac12\ln^2(1-\alpha-\beta)+\frac12\ln^2\bar\alpha+\frac12\ln^2\bar\beta-\ln\alpha\ln\bar\alpha-\ln\beta\ln\bar\beta
\notag\\
&\quad-\frac12\ln\alpha-\frac12\ln\beta
+\frac{\bar\alpha}\alpha\ln\bar\alpha+\frac{\bar\beta}\beta\ln\bar\beta\Big]\,.
\end{align}
The corresponding one- and two-loop anomalous dimensions
\begin{align}
   \mathbb{H}^{(\ell)} (z_1-z_2)^N = \gamma_N^{(\ell)} (z_1-z_2)^N\,, \qquad \ell=1,2
\end{align}
are well known and can be found, e.g., in Ref.~\cite{Moch:2004pa}.

Next, the generator of special conformal transformations reads
\begin{align}
S_+(a) &= S_+^{(0)} +a\, \Delta S_+^{(1)}+ a^2\, \Delta S_+^{(2)}+\ldots
\end{align}
where~\eqref{S+loop}
\begin{align}
 \Delta S_+^{(1)} &=
 (z_1+z_2) \left(\beta_{0} + \frac12 \mathbb{H}^{(1)}\right)
 + z_{12} \Delta^{(1)}_+\,,
\notag\\
 \Delta S_+^{(2)} &=
(z_1+z_2) \left(\beta_{1} + \frac12 \mathbb{H}^{(2)}\right) + \frac1{4}\big[\mathbb{H}^{(2)},z_1+z_2\big]
 + z_{12} \Delta^{(2)}_+\,.
\label{S+loop12}
\end{align}
The one-loop ``conformal anomaly'' contribution, $\Delta_+^{(1)}$, is very
simple~\cite{Belitsky:1998vj,Braun:2014vba}
\begin{align}
\Delta^{(1)}_+f(z_1,z_2)
&= 2C_F\int_0^1d\alpha \int_0^1 du \frac{\bar\alpha}\alpha
\Big[f(z_{12}^{\alpha u},z_2)-f(z_1,z_{21}^{\alpha u})\Big]\,
\notag\\&=
- 2C_F\int_0^1d\alpha\, \left(\frac{\bar\alpha}\alpha+\ln\alpha\right)
\Big[f(z_{12}^{\alpha},z_2)-f(z_1,z_{21}^{\alpha})\Big]\,.
\label{Delta+oneloop}
\end{align}
The expression for the two-loop anomaly, $\Delta_+^{(2)}$, represents our main result.
It can be written as
\begin{align}
  \Delta_+^{(2)} f(z_1,z_2) &=
\int_0^1\!d\alpha\!\int_0^{\bar\alpha} d\beta \Big[\omega(\alpha,\beta) + \omega^{\mathbb{P}}(\alpha,\beta) \mathbb{P}_{12}\Big]
\Big[f(z_{12}^{\alpha},z_{21}^\beta)-f(z_{12}^{\beta},z_{21}^\alpha)\Big]
\nonumber\\
&\quad
+ \int_0^1\!du\int_0^1\!dt \, \varkappa(t)\,\Big[f(z_{12}^{ut},z_2) - f(z_1,  z_{21}^{ut})\Big].
\label{Delta+twoloops}
\end{align}
 The kernels $\varkappa(t)$, $\omega(\alpha, \beta)$, $\omega^{\mathbb{P}}(\alpha,\beta)$ receive contributions of three
different color structures
\begin{align}
\varkappa(t) &= C_F^2\, \varkappa_{FF}(t)+ C_F C_A \,\varkappa_{FA}(t)+C_F\beta_0 \varkappa_{bF}(t)\,,
\notag    \\
\omega(\alpha,\beta) &= C_F^2\, \omega_{FF}(\alpha,\beta)+ C_F C_A \,\omega_{FA}(\alpha,\beta)\,,
\notag    \\
\omega^{\mathbb{P}}(\alpha,\beta) &= C_F^2 \, \omega_{FF}^{\mathbb{P}}(\alpha,\beta)+ C_F C_A \, \omega_{FA}^{\mathbb{P}}(\alpha,\beta)\,.
\end{align}
Alternatively one can write the results separating the contributions of planar diagrams
and the non-planar $1/N_c$ suppressed corrections
\begin{align}
\varkappa(t) & =C_F^2\, \varkappa_{\text{P}}(t)+ \frac{C_F}{N_C} \,\varkappa_{FA}(t)+C_F\beta_0 \varkappa_{bF}(t),
\notag   \\
\omega(\alpha,\beta) &= C_F^2\, \omega_{\text{P}}(\alpha,\beta)+\frac{C_F}{N_C} \omega_{FA}(\alpha,\beta),
\notag   \\
\omega^{\mathbb{P}}(\alpha,\beta) &=\frac{C_F}{N_C} \omega_{FA}^{\mathbb{P}}(\alpha,\beta),
\end{align}
where, obviously, $\varkappa_{\text{P}} =  \varkappa_{FF} + 2\varkappa_{FA} $ etc.,
and we took into account that the terms involving quark permutations on the light cone
do not receive planar contributions, $\omega_{FA}^{\mathbb{P}}=-\frac12\omega_{FF}^{\mathbb{P}}$.

Note that the term $\sim \beta_0$ in $\varkappa(t)$ arises by choice, rewriting the contribution
proportional to the number of quark flavors $N_f$ in terms of $\beta_0$. This rewriting is not mandatory and
is only motivated in the present context by resulting in somewhat simpler expressions.
In contrast, the terms $(z_1+z_2)\beta_{\ell-1}$ in the expression for $\Delta S^{(\ell)}_+$ involve
the ``genuine'' QCD beta-function.

Explicit expressions for the ``two-particle'' kernels $\omega, \omega^{\mathbb{P}}$ are:
\begin{align}
\omega_{FF}(\alpha,\beta) &= 4\biggl[
\left(\alpha-\frac1\alpha\right)\Big[\Li_2\left(\frac \beta{\bar\alpha}\right)-\Li_2(\beta)-\Li_2(\alpha)-\frac14\ln^2\bar\alpha\Big] -\alpha \Big[\Li_2(\alpha)-\Li_2(1)\Big]
\notag\\
&\quad
- \frac{\alpha+\beta}2\ln\alpha\ln\bar\alpha
+\frac14\left(\beta\ln^2\bar\alpha  -\alpha\ln^2\alpha \right)
-\frac\alpha\tau\Big(\tau\ln\tau+\bar\tau \ln\bar\tau\Big)
\notag\\
&\quad
+\frac14\left(\beta-2\bar\alpha+\frac{2\beta}\alpha\right)\ln\bar\alpha
+
\frac12\left(\bar\alpha-\frac\alpha{\bar\alpha}-3\beta\right)\ln\alpha-\frac{15}4\alpha
\biggr],
\notag\\
\omega_{FA}(\alpha,\beta) &
=2\biggl[\left(\frac1\alpha-\alpha\right)\Big[\Li_2\left(\frac\beta{\bar\alpha}\right)-\Li_2(\beta)-2\Li_2(\alpha)-\ln\alpha\ln\bar\alpha\Big]
+\frac\alpha{\tau}\Big(\tau\ln\tau+\bar\tau \ln\bar\tau\Big)
\notag\\
&\quad
-\bar\beta\ln\alpha-\frac{\bar\alpha}{\alpha}\ln\bar\alpha\biggr],
\notag\\
\omega_{FA}^{\mathbb{P}}(\alpha,\beta) &=
2  \biggl[
\left(\bar\alpha-\frac1{\bar\alpha}\right)\Big[\Li_2\left(\frac\alpha{\bar\beta}\right)-\Li_2(\alpha)-\ln\bar\alpha\ln\bar\beta\Big]
+\alpha \bar\tau \ln\bar\tau
+\frac{\beta^2}{\bar\beta}\ln\bar\alpha
\biggr]
\end{align}
and for the planar combination $\omega_{\text{P}} = \omega_{FF} + 2 \omega_{FA}$
\begin{align}
 \omega_{\text{P}}(\alpha,\beta)&
 = \frac4\alpha\Big[\Li_2(\bar\alpha)-\Li_2(1)\Big]
+ \frac1{\alpha} \ln^2\bar\alpha-(\alpha-\beta)\ln^2\left(\frac{\alpha}{\bar\alpha}\right)-\beta\ln^2\alpha
\notag\\
&\quad
+2\alpha\left(\frac{\pi^2}3-\frac{15}2\right)-2\left(\alpha+\beta+ \frac1{\bar\alpha}\right)\ln\alpha
+\Big(\beta-2\bar\alpha\Big)\left(1+\frac2\alpha\right)\ln\bar\alpha\,.
\end{align}
For the ``one-particle'' kernels $\varkappa(t)$ we obtain
\begin{align}
\varkappa_{bF}(t) &=
 - 2 \frac{\bar t}{t}
 \Big(  \ln \bar t   + \frac53  \Big),
%\notag\\
%\varkappa_{FA}(t) & =\frac{2\bar t}{t} (2+t)\Big[\Li_2(\bar t)-\Li_2(t)\Big]
%-2\frac{2- t}t\Big[t\ln t+{\bar t}\ln\bar t\Big]
%%\notag\\
%%&\quad
% - \frac{\pi^2} 3 \bar t  -\frac83\frac{\bar t}t-(1-2t)\,,
\notag\\
\varkappa_{FA}(t) & =\frac{2\bar t}{t}\biggl\{ (2+t)\Big[\Li_2(\bar t)-\Li_2(t)\Big]
-(2- t)\Big(\frac{t}{\bar t}\ln t+\ln\bar t\Big)
 - \frac{\pi^2}6 t  -\frac43-\frac{t}{2}\left(1-\frac t{\bar t}\right)\biggr\},
\notag\\
\varkappa_{\text{P}}(t) & =4\bar t \Big[\Li_2(\bar t)-\Li_2(1)\Big]
+4\left(\frac{t^2}{\bar t}-\frac{2\bar t}t\right)\Big[\Li_2(t)-\Li_2(1)\Big]
-2t\ln t\ln\bar t
-\frac{\bar t}t (2-t) \ln^2\bar t
\notag\\
&\quad
+\frac{t^2}{\bar t}\ln^2 t-2\left(1+\frac1t\right)\ln\bar t
-2\left(1+\frac1{\bar t}\right)\ln t - \frac{16}3\frac{\bar t}t -1-5t\,.
\end{align}
The last expression can also be rewritten as
\begin{align}
\varkappa_{\text{P}}(t) & =-4\bar t \Li_2(1)
+4\left(\frac1{\bar t}-\frac2t\right)\Big[\Li_2(t)-\Li_2(1)\Big]
-2(2-t)\ln t\ln\bar t
-\frac{\bar t}t (2-t) \ln^2\bar t
\notag\\
&\quad
+\frac{t^2}{\bar t}\ln^2 t-2\left(1+\frac1t\right)\ln\bar t
-2\left(1+\frac1{\bar t}\right)\ln t - \frac{16}3\frac{\bar t}t -1-5t\,.
\end{align}
The result for $\varkappa_{FF}(t)$ can easily be obtained by subtracting
$\varkappa_{\text{P}}(t) - 2\varkappa_{FA}(t)$.
Note that we prefer to write the corresponding contribution to $\Delta_+$ (second line in \eqref{Delta+twoloops}) as a nested
integral in auxiliary $u,t$ variables. One of these integrations can be taken trivially (cf.~\eqref{Delta+oneloop})
resulting in somewhat more complicated expressions involving $\Li_3$-functions.

Finally, the commutator term $\frac1{4}\big[\mathbb{H}^{(2)},z_1+z_2\big]$ in Eq.~\eqref{S+loop12}
can be written as
\begin{multline}
 \frac1{4}\big[\mathbb{H}^{(2)},z_1+z_2\big] f(z_1,z_2) =
(z_1-z_2)\biggl\{\int_0^1 d\alpha\,\bar\alpha\, h(\alpha)\,\Big[f(z_{12}^\alpha,z_2)-f(z_1,z_{21}^\alpha)\Big]
\\
-\int_0^1d\alpha\int_0^{\bar\alpha}d\beta \,\alpha \Big(\chi(\alpha,\beta)-\chi^{\mathbb{P}}(\alpha,\beta)\mathbb{P}_{12}\Big)\,
\Big[f(z_{12}^\alpha,z_{21}^\beta)-f(z_{12}^\beta,z_{21}^\alpha)\Big]\biggr\}.
\end{multline}
This term can be added to the result for  $\Delta^{(2)}_+$~\eqref{Delta+twoloops} but keeping it separate seems to
be more convenient for applications.

%
%%%%%%%%%%%%%%%%%%%%%%%%%%%%%%%%%%%%%%%%%%%%%%%%%%
%

%%%%%%%%%%%%%%%%%%%%%%%%%%%%%%%%%%%%%%%%%%%%%%%%%%%%%%%%%%%%%%%%%%%%%%%%%%%%%%%%%%%%%%%%%%%%%%%%%%%%%%%%%%%%%%%%%%%%%%
\section{Conclusions}\label{sec:Conclusions}
%%%%%%%%%%%%%%%%%%%%%%%%%%%%%%%%%%%%%%%%%%%%%%%%%%%%%%%%%%%%%%%%%%%%%%%%%%%%%%%%%%%%%%%%%%%%%%%%%%%%%%%%%%%%%%%%%%%%%%

QCD evolution equations in minimal subtraction schemes have a hidden symmetry:
One can construct three operators that commute with the evolution kernel and form an $SL(2)$ algebra,
i.e. they satisfy (exactly) the $SL(2)$ commutation relations.
In this paper we find explicit expressions for these operators to two-loop accuracy.
On this way we make a digression to the $4-2\epsilon$ dimensional world, where conformal symmetry of QCD
is restored on quantum level at the specially chosen (critical) value of the coupling, and at the same time
the theory is regularized allowing one to use the standard renormalization procedure for the relevant
Feynman diagrams. We want to emphasize that the procedure is valid to all orders in perturbation theory
and the result obtained in this way is complete, i.e. it includes automatically all terms that can be identified
as due to a nonvanishing QCD $\beta$-function (in the physical theory in four dimensions).
To avoid misunderstanding, we stress that QCD in $d=4$ dimensions is
certainly not a conformal theory. The symmetry that we uncover is the symmetry of \emph{RG equations} in QCD
in a specially chosen (dimensional) regularization scheme.

It is well known that (QCD) has a non-trivial fixed point in strictly four space-time dimensions for a
range of values of the number of quark flavors $9\le N_f \le 16$ known as the conformal window,
and in the last years there has been increasing interest in the study of the phase structure of such theories
(Banks-Zaks fixed point \cite{Banks:1981nn}) on the lattice, see e.g.~\cite{DeGrand:2015zxa} for a recent review.
  Our result for the generators of conformal transformations,
where one has to use the appropriate values of $N_f$ and $N_c$, is valid for QCD within the conformal window as well,
to the $\mathcal{O}(a^2_\ast)$ accuracy in the critical coupling.

The main motivation for this study is to obtain three-loop evolution equations for the
generalized hadron parton distributions and light-cone meson distribution amplitudes that are relevant for
the large-scale experimental studies of hard exclusive reactions in the coming decade.
The present work presents the first step in this direction. The remaining calculation
can be done in several ways. One possibility is to solve the system of linear differential
equations~\eqref{nest2}, as demonstrated in Ref.~\cite{Braun:2014vba} to the two-loop accuracy.
Alternatively, one can exploit the well-known observation \cite{Bukhvostov:1985rn} that
the evolution kernel must be a function of the quadratic Casimir operator of the collinear conformal
group. This function can be found from the spectrum of anomalous dimensions.
Yet another possibility is to bypass the explicit construction of the kernel and try to find
directly the solutions (conformal operators) by constructing a unitary transformation $U$ that brings
the generator $S_+$ to its canonical form, $U S_+ U^\dagger = S_+^{(0)}$. Utility of each of these methods requires a separate
study that goes beyond the scope of this work.

Last but not least, the explicit perturbative construction of the generators of conformal transformations can
be interesting in the context of the AdS/CFT correspondence for the maximally supersymmetric $N=4$ Yang-Mills
theory. The general structure of this expansion should be similar to what is obtained in this
paper for QCD, but the answer is expected to be simpler. It would be interesting to do this calculation and compare the result
with the algebraic approach in Ref.~\cite{Zwiebel:2005er}.

%%%%%%%%%%%%%%%%%%%%%%%%%%%%%%%%%%%%%%%%%%%%%%%%%%

%\vspace*{1cm}
%\begin{center}
%{\Huge\bf \Coffeecup}
%\end{center}
%\vspace*{1cm}

%%%%%%%%%%%%%%%%%%%%%%%%%%%%%%%%%%%%%%%%%%%%%%%%%%
%

%%%%%%%%%%%%%%%%%%%%%%%%%%%%%%%%%%%%%%%%%%%%%%%%%%%%%%%%%%%%%%%%%%%%%%%%%%%%%%%%%%%%%%%%%%%%%%%%%%%%%%%%%%%%%%%%%%%%%%
\section*{Acknowledgments}
%%%%%%%%%%%%%%%%%%%%%%%%%%%%%%%%%%%%%%%%%%%%%%%%%%%%%%%%%%%%%%%%%%%%%%%%%%%%%%%%%%%%%%%%%%%%%%%%%%%%%%%%%%%%%%%%%%%%%%
\addcontentsline{toc}{section}{Acknowledgments}

This study was supported by Deutsche Forschungsgemeinschaft (DFG) with the grant MO~1801/1-1.

\appendix

%%%%%%%%%%%%%%%%%%%%%%%%%%%%%%%%%%%%%%%%%%%%%%%%%%%%%%%%%%%%%%%%%%%%%%%%%%%%%%%%%%%%%%%%%%%%%%%%%%%%%%%%%%%%%%%%%%%%%%
%%%%%%%%%%%%%%%%%%%%%%%%%%%%%%%%%%%%%%%%%%%%%%%%%%%%%%%%%%%%%%%%%%%%%%%%%%%%%%%%%%%%%%%%%%%%%%%%%%%%%%%%%%%%%%%%%%%%%%
%%%%%%%%%%%%%%%%%%%%%%%%%%%%%%%%%%%%%%%%%%%%%%%%%%%%%%%%%%%%%%%%%%%%%%%%%%%%%%%%%%%%%%%%%%%%%%%%%%%%%%%%%%%%%%%%%%%%%%
%%%%%%%%%%%%%%%%%%%%%%%%%%%%%%%%%%%%%%%%%%%%%%%%%%%%%%%%%%%%%%%%%%%%%%%%%%%%%%%%%%%%%%%%%%%%%%%%%%%%%%%%%%%%%%%%%%%%%%

\section*{Appendices}
\addcontentsline{toc}{section}{Appendices}

\renewcommand{\theequation}{\Alph{section}.\arabic{equation}}
\renewcommand{\thetable}{\Alph{table}}
\setcounter{section}{0}
\setcounter{table}{0}

%%%%%%%%%%%%%%%%%%%%%%%%%%%%%%%%%%%%%%%%%%%%%%%%%%%%%%%%%%%%%%%%%%%%%%%%%%%%%%%%%%%%%%%%%%%%%%%%%%%%%%%%%%%%%%%%%%%%%%
\section{BRST transformations}\label{app:BRST}
%%%%%%%%%%%%%%%%%%%%%%%%%%%%%%%%%%%%%%%%%%%%%%%%%%%%%%%%%%%%%%%%%%%%%%%%%%%%%%%%%%%%%%%%%%%%%%%%%%%%%%%%%%%%%%%%%%%%%%

The QCD action~\eqref{SQCD} is invariant under the BRST transformations~\cite{Becchi:1975nq}
\begin{align}
\delta q & =ig t^a q c^a\delta\lambda\,
%&&
%\delta \bar q=-ig \bar q t^a  c^a\delta\lambda
&
\delta A_\mu &=\left(\partial_\mu c^a+g f^{abc}A^b_\mu c^c \right)\delta\lambda\,,
\notag\\
\delta c^a & =\frac12 g f^{abc} c^b c^c\delta\lambda\,
&
\delta \bar c^a & =-\frac1\xi (\partial A^a)\delta \lambda\,.
\end{align}
Transformation rules for the renormalized fields are obtained by replacing in the above equations
$\Phi\mapsto \Phi_0$, $g \to g_0$, $\xi\to \xi_0$, $\delta \lambda\to \delta\lambda_0$
and writing the bare fields and couplings in terms of the renormalized ones:
$\Phi_0=Z_\Phi \Phi$, $g_0 = Z_g g$, $\xi_0 = Z_\xi \xi$. The renormalized BRST transformation parameter
$\delta\lambda$ is defined as $\delta\lambda_0=Z_c Z_A \delta\lambda$ so that the last equation
has the same form for bare and renormalized quantities,
\begin{align}
\delta \bar c^a=-\frac1\xi (\partial A^a)\delta \lambda\,.
\end{align}
There are two BRST operators which appear in our analysis. One of them is
$\mathcal{B}_\mu$~\eqref{NBmu}, which is the BRST variation of $\bar c^a A^a_\mu$~\cite{Belitsky:1998gc}
\begin{align}
\mathcal{B}_\mu(x)= Z_c^2\bar c D^\mu c-\frac1\xi A^\mu (\partial A) =\frac{\delta}{\delta\lambda_R}\bar c^a A^a_\mu\,,
\end{align}
%
%It can be shown that this is a finite operator, i.e. $[\mathcal{B}_\mu]=\mathcal{B}_\mu$.
and another one is
\begin{align}\label{barcEOM}
\mathcal{B}=\frac{\delta}{\delta\lambda_R}\bar c^a (\partial A^a)=-\frac1\xi (\partial A)^2+ Z_c^2 \bar c \partial D c\,.
\end{align}
Thanks to this identity the gauge fixing term in the action can be represented as a sum of the EOM and BRST exact operators
\begin{align}\label{fixing-B}
\frac1\xi (\partial A)^2=-\mathcal{B}+\Omega_{\bar c}\,, \qquad\quad
\Omega_{\bar c} = Z_c^2 \bar c \partial D c = \bar c(x) \frac{\delta S_R}{\delta\bar c(x)}\,.
\end{align}
One can show that $\mathcal{B}_\mu$ is a finite operator, i.e. $[\mathcal{B}_\mu]=\mathcal{B}_\mu$, while $\mathcal{B}$
is not, $[\mathcal{B}]=Z_B \mathcal{B}+Z_{B^\mu} \partial^\mu \mathcal{B}_\mu +~\text{EOM}$.

%%%%%%%%%%%%%%%%%%%%%%%%%%%%%%%%%%%%%%%%%%%%%%%%%%%%%%%%%%%%%%%%%%%%%%%%%%%%%%%%%%%%%%%%%%%%%%%%%%%%%%%%%%%%%%%%%%%%%%
\section{Renormalization group analysis}\label{App:action}
%%%%%%%%%%%%%%%%%%%%%%%%%%%%%%%%%%%%%%%%%%%%%%%%%%%%%%%%%%%%%%%%%%%%%%%%%%%%%%%%%%%%%%%%%%%%%%%%%%%%%%%%%%%%%%%%%%%%%%

A generic gauge invariant operator can mix under renormalization with: A)~gauge invariant operators,
$\mathcal{O}_A$,  B)~BRST exact operators, $\mathcal{O}_B=\delta_{BRST} \widetilde{\mathcal{O}}_B$,  and
C)~Equation of motion (EOM) operators,
$\mathcal{O}_C$,  see, e.g., Ref.~\cite{Collins}. Schematically,
\begin{align}
[\mathcal{O}_K]=Z_{KM} \mathcal{O}_{M}, && \text{where\,\,\,} K,M=A,B,C\,.
\end{align}
Importantly, the matrix $Z_{KK'}$ has an upper triangular form. Thus
\begin{align}
[\mathcal{O}_A]=Z_{AA} \mathcal{O}_{A}+Z_{AB} \mathcal{O}_{B}+Z_{AC} \mathcal{O}_{C}\,.
\end{align}
The last two terms do not contribute, however, to the correlation function of two gauge
invariant operators at different space-time points $x\neq y$,
\begin{align}
\VEV{[\mathcal{O}_A](x)[\mathcal{O}_{A'}](y)}=Z_{AA}  Z_{A'A'}\VEV{\mathcal{O}_{A}(x)\mathcal{O}_{A'}(y)}\,.
\end{align}
Indeed, class $C$, EOM operators, give rise to contact terms $\sim \delta^{(d)}(x-y)$,
whereas operators of class $B$ do not contribute to the correlation function due to the BRST invariance of the QCD action.

In this work we consider flavor-nonsinglet leading twist operators made of the quark and antiquark field and
covariant derivatives.  For these operators any counterterms  (of any class) contain the same pair of quark fields,
$\bar q,\, q$. A simple dimensional analysis shows, however, that possible operators of class $B$ and $C$  cannot,
in this case, be traceless in all Lorentz indices --- so that they cannot appear as counterterms in leading-twist operators.
Thus only gauge-invariant operators can contribute, $[\mathcal{O}_A]=Z_{AA} \mathcal{O}_{A}$.

The scale and conformal variation of the action~\eqref{DC} contains the ``symmetry breaking'' operator
$\mathcal{N}$ \eqref{NBmu}.  Making use of Eq.~\eqref{fixing-B} we can write it in the form
\begin{align}\label{NFB}
\mathcal{N}=\epsilon \Big(\frac12 Z_A^2 F^2- \mathcal{B}+\Omega_{\bar c}\Big)\,.
\end{align}
Note that the last two terms drop out from the correlation functions of gauge invariant operators and
$\mathcal{N}$.

Our goal is to express the operator $\mathcal{N}$ in terms of renormalized operators. From
the general operator mixing pattern discussed above we expect the following structure:
\begin{align}
[\mathcal{B}]&= Z_B \mathcal{B} + Z_{B^\mu}\partial_\mu \mathcal{B}^\mu+\sum_\Phi Z_{B\Phi}\Omega_\Phi\,,
\notag\\
[F^2]&=Z_F F^2+Z_{FB} \mathcal{B}+Z_{FB^\mu}\partial_\mu \mathcal{B}^\mu + \sum_\Phi Z_{F\Phi}\Omega_\Phi\,,
\end{align}
where $\Phi = \{q,\bar q, A, c,\bar c\}$ and $\Omega_\Phi=\Phi {\delta S_R}/{\delta\Phi}$ are EOM operators.
One can re-express the operators $F^2$ and $\mathcal{B}$ appearing in~\eqref{NFB} in terms of the corresponding
renormalized operators with, in general, singular coefficients, $\mathcal{B}=Z_b^{-1} [\mathcal{B}]+\ldots$.
 Taking into account that $[\Omega_\Phi]=\Omega_\Phi$ and $[\mathcal{B}^\mu]=\mathcal{B}^\mu$ we obtain
\begin{align}\label{Nren}
\mathcal{N}=\epsilon \biggl(\frac12 \widetilde Z_F[F^2]-\widetilde Z_{B}[\mathcal{B}]+\widetilde Z_{B^\mu}\partial_\mu[\mathcal{B}^\mu]
+\sum_\Phi \widetilde Z_\Phi\, \Omega_\Phi
\biggr).
\end{align}
Next, consider for a moment the Ward Identities of the type \eqref{SCWI} for the products of (renormalized) fields $\Phi_k$.
Since the terms involving the variations of the fields are finite (they reduce to a differential operator
applied to a finite Green function, cf.~\eqref{scaleWI}), the term with the variation of the action must be finite as well;
hence the operator $\mathcal{N}$ is also finite (up to possible terms containing two total derivatives)
and, therefore, the product $\epsilon  \widetilde Z_K$ is finite for all factors $\widetilde Z_K$ appearing in~\eqref{Nren}.
Taking into account that $\widetilde Z_{F},\widetilde Z_{B} =1+O(1/\epsilon)$, whereas all
other factors only contain poles, $Z_{\mathcal{B}^\mu},Z_{\Phi}=O(1/\epsilon)$, we get
\begin{align}\label{Nrc}
\mathcal{N}=\frac12(\epsilon+r_F) [F^2]-(\epsilon+r_B) [\mathcal{B}]+r_{B^\mu}\partial_\mu[\mathcal{B}^\mu]
-\sum_\Phi r_\Phi\, \Omega_\Phi+\epsilon\, \Omega_{\bar c}\,,
\end{align}
where the coefficients $r_K$ do not depend on $\epsilon$; they are functions of the coupling constant and the
gauge fixing parameter.

These coefficients for the operators that do not involve total derivatives (alias whose matrix elements do not
vanish for zero momenta) can be fixed from the study of the differential vertex operator insertions, see below.
Note that $\Omega_{\bar q}-\Omega_{q}=\partial^\mu \bar q\gamma_\mu q$ and
$\Omega_{\bar c}-\Omega_{c}=  \partial^\mu[\bar c {D}_\mu c - \partial_\mu\bar c c]= \partial^\mu{\Omega}_\mu$
do reduce to total derivatives, so that we can determine the sum of the
coefficients, $r_q+r_{\bar q}$ and $r_{c}+r_{\bar c}$, but not their difference.
Invoking charge symmetry arguments one can argue that $r_q-r_{\bar q}=0$, whereas $r_c-r_{\bar c}$
and the coefficient $r_{B^\mu}$ cannot be determined in this approach.

As well known, derivatives of Green functions of fundamental fields with respect to the couplings
give rise to zero momentum insertions of the differential vertex operators
\begin{align}
 g\partial_g \big\langle \Phi_1(x_1)\ldots \Phi_n(x_n)\big\rangle &=
\int d^dx\, \big\langle g\partial_g\mathcal{L}_R(x)\Phi_1(x_1)\ldots \Phi_n(x_n)\big\rangle
\end{align}
and similarly for $g\partial_g \to \xi\partial_\xi$. Since the expression on the l.h.s. is finite,
one concludes that the correlation function with an insertion of
$g\partial_g\mathcal{L}_R$ or $\xi\partial_\xi\mathcal{L}_R$ at
zero momentum is also finite. Thus both $g\partial_g\mathcal{L}_R$ and $\xi\partial_\xi\mathcal{L}_R$
(up to total derivatives) can be written as a sum of renormalized operators with finite coefficients.

For further analysis it is convenient to redefine, temporary,
the bare gluon field $A_0 \mapsto G_0= g_0 A_0$ so that
\begin{align}
\mathcal{L}_R=\frac1{g_0^2}\left(\frac14 F_0^2+\frac1{2\xi_0} (\partial G_0)^2\right)+\bar q_0 i \slashed{D}q_0+\bar c_0 \partial D c_0\,,
\end{align}
and the last two terms do not depend on $g_0$ and $\xi_0$. Let us evaluate the derivatives
\begin{align}\label{}
\partial_g \mathcal{L}_R=&\partial_g \mathcal{L}(\Phi_0,g_0,\xi_0)=
\frac{\delta \mathcal{L}}{\delta \Phi_0}\frac{\partial\Phi_0}{\partial g}
+\frac{\partial \mathcal{L}}{\partial g_0}\frac{\partial g_0}{\partial g}+
\frac{\partial \mathcal{L}}{\partial \xi_0}\frac{\partial \xi_0}{\partial g}\,,
\notag\\
\partial_\xi \mathcal{L}_R=&\partial_\xi \mathcal{L}(\Phi_0,g_0,\xi_0)=
\frac{\delta \mathcal{L}}{\delta \Phi_0}\frac{\partial\Phi_0}{\partial \xi}
+\frac{\partial \mathcal{L}}{\partial g_0}\frac{\partial g_0}{\partial \xi}
+\frac{\partial \mathcal{L}}{\partial \xi_0}\frac{\partial \xi_0}{\partial \xi}\,.
\end{align}
Taking into account that $\partial_g \Phi_0= \Phi_0\partial_g \ln Z_\Phi $,
in particular for the redefined gluon field $\partial_g G_0= G_0\,\partial_g\ln (g Z_g Z_A )$,
and also
\begin{align}
g_0\partial_{g_0}{ \mathcal{L}_R}=-\frac2{ g_0^2}\left(\frac1{4} F_0^2+\frac1{2\xi_0 }(\partial G_0)^2\right), &&
\xi_0\partial_{\xi_0}{ \mathcal{L}_R}=-\frac1{2g_0^2\xi_0 }(\partial G_0)^2
\end{align}
together with
\begin{align}
\partial_{g} g_0=-\epsilon g_0/\beta(g)\,,&& \partial_{g} \xi_0=2\xi_0 \partial_g \ln Z_A\,, &&
\partial_{\xi} \xi_0=\xi_0\big(1+ 2\partial_\xi \ln Z_A\big)\,,
\end{align}
one obtains
\begin{align}\label{gxiS}
\partial_g \mathcal{L}_R & =\frac{2\epsilon}{\beta(g)}\left( \mathcal{L}_R^{YM}+\mathcal{L}_R^{gf}\right)+\Omega_A \partial_g\ln(gZ_gZ_A)+
\sum_{\Phi\neq A}  \Omega_{\Phi}\,\partial_g \ln Z_\Phi
- \frac1{\xi}(\partial A)^2\partial_g\ln Z_A\,,
\notag\\
\xi\partial_\xi \mathcal{L}_R & =-\frac1{2\xi}(\partial A)^2 (1+2\xi\partial_\xi\ln  Z_A)+
\sum_{\Phi} \Omega_\Phi\, \xi \partial_\xi \ln Z_\Phi\,,
\end{align}
where $\mathcal{L}_R^{YM}$ and $\mathcal{L}_R^{gf}$ are the gauge (Yang-Mills) and the gauge-fixing parts of the
(renormalized) QCD Lagrangian.

The expressions on the r.h.s. of the two equations in~\eqref{gxiS} have the following structure:
\begin{align}
g\partial_g \mathcal{L}_R  =-2\left(\mathcal{L}^{YM}+\mathcal{L}^{gf}-\frac12 A\frac{\delta \mathcal{L}}{\delta A}\right)+\ldots\,,
%\text{series in}\ \  \frac1\epsilon
&&
\xi\partial_\xi \mathcal{L}_R  =-\frac1{2\xi}(\partial A)^2+\ldots %\text{series in}\ \  \frac1\epsilon
%\sum_{\Phi}  \Phi \frac{\delta S_R}{\delta \Phi} \xi \partial_\xi \ln Z_\Phi
\end{align}
where the ellipses stand for a series in $1/\epsilon$. Since the operators on the l.h.s. are finite,
the addition of these terms (ellipses) effectively amounts to a subtraction of divergences so that
the sum is nothing but, by definition, a renormalized operator in MS scheme. Thus
\begin{align}
g\partial_g \mathcal{L}_R  &=-2\left[\mathcal{L}^{YM}+\mathcal{L}^{gf}-\frac12 \Omega_A\right]
= -2\left[\mathcal{L}^{YM}+\mathcal{L}^{gf}\right] +  \Omega_A \,,
\notag\\
\xi\partial_\xi \mathcal{L}_R  &=-\frac1{2\xi}[\partial A)^2]\,.
\end{align}
Replacing $\partial_g \mathcal{L}_R$, $\xi\partial_\xi \mathcal{L}_R$ on the l.h.s. of Eqs.~\eqref{gxiS} by these
expressions one obtains, after a little rewriting
\begin{align}
\label{eq:B1}
 2\epsilon \left( \mathcal{L}_R^{YM}+\mathcal{L}_R^{gf}\right) &=
-\frac{2\beta(g)}{g}\left[\mathcal{L}^{YM}+\mathcal{L}^{gf}\right]
 -\Omega_A \mathcal{D}_g\ln(Z_gZ_A)-
\sum_{\Phi\neq A}  \Omega_{\Phi}\,\mathcal{D}_g \ln Z_\Phi
\notag\\&{}\hspace*{4mm}
+ \frac1{\xi}(\partial A)^2\mathcal{D}_g\ln Z_A\,,
\end{align}
where $\mathcal{D}_g = \beta(g)\partial_g$, and also
\begin{align}\label{N2}
-\frac1{2\xi}[(\partial A)^2] & =-\frac1{2\xi}(\partial A)^2 (1+2\xi\partial_\xi\ln  Z_A)+
\sum_{\Phi} \Omega_\Phi\, \xi \partial_\xi \ln Z_\Phi\,.
\end{align}
We remind that these results are valid for zero-momentum insertions, or, equivalently,
upon integration $\int d^dx$ over all space-time points.

Finally, note that  $\gamma_\Phi=M\partial_M \ln Z_\Phi=(\beta_g \partial_g+\beta_\xi
\partial_\xi)\ln Z_\Phi$ and $\beta_\xi=-2\xi \gamma_A$ so that $\mathcal{D}_g\ln Z_A=\gamma_A (1+2\xi\partial_\xi \ln Z_A)$.
Thus the last term in \eqref{eq:B1} can be rewritten as
$-\frac1{2\xi}[(\partial A)^2] - \sum_{\Phi} \Omega_\Phi\, \xi \partial_\xi \ln Z_\Phi$ and collecting all contributions
we obtain
\begin{align}\label{NN}
&\mathcal{N}(x)  =-\frac{2\beta(g)}{g}\left[\mathcal{L}^{YM}+\mathcal{L}^{gf}%-\frac12 \Omega_A
\right]-(\gamma_A+\gamma_g)\Omega_A-
\sum_{\Phi\neq A} \gamma_\Phi \Omega_{\Phi}
+ \frac{\gamma_A}{\xi}[(\partial A)^2]+\ldots
\end{align}
where the ellipses stand for total derivative operators. From this expression one can read
the results for the coefficients $r_K$ defined in Eq.~\eqref{Nrc}:
\begin{align}
r_F=\gamma_g\,,&& r_B=r_A=\gamma_g+\gamma_A\,, &&r_q=r_{\bar q}=\gamma_q\,, &&r_c+r_{\bar c}=2\gamma_c+\gamma_g+\gamma_A\,.
\end{align}
To avoid misunderstanding note that in the derivation we did not use criticality so that
the result in  Eq.~\eqref{NN} is valid for arbitrary coupling.

%%%%%%%%%%%%%%%%%%%%%%%%%%%%%%%%%%%%%%%%%%%%%%%%%%%%%%%%%%%%%%%%%%%%%%%%%%%%%%%%%%%%%%%%%%
\section{Results for separate diagrams in Feynman gauge}\label{App:diagrams}
%%%%%%%%%%%%%%%%%%%%%%%%%%%%%%%%%%%%%%%%%%%%%%%%%%%%%%%%%%%%%%%%%%%%%%%%%%%%%%%%%%%%%%%%%%

%%%%%%%%%%%%%%%%%%%%%%%%%%%%%%%%%%%%%%%%%%%%%%%%%%%%%%%%%%%%%%%%%%%%%%%%%%%%%%%%%%%%%%%%%%
\subsection{Evolution kernel}
%%%%%%%%%%%%%%%%%%%%%%%%%%%%%%%%%%%%%%%%%%%%%%%%%%%%%%%%%%%%%%%%%%%%%%%%%%%%%%%%%%%%%%%%%%

The contributions to the evolution kernel from the diagrams in Fig.~\ref{fig:Diagrams}(a)--(p)
(including symmetric diagrams with the interchange of the quark and the antiquark)
can be written in the following form:
\begin{eqnarray}
  [\mathbb{H} \mathcal{O}](z_1z_2) &=&
-4 \int_0^1\!d\alpha\!\int_0^{\bar\alpha} d\beta \Big[\chi(\alpha,\beta) + \chi^{\mathbb{P}}(\alpha,\beta) \mathbb{P}_{12}\Big]
\Big[\mathcal{O}(z_{12}^{\alpha},z_{21}^\beta)+\mathcal{O}(z_{12}^{\beta},z_{21}^\alpha)\Big]
\nonumber\\&&{}
 -4 \int_0^1\!du \, h (u) \Big[2 \mathcal{O}(z_1,z_2) - \mathcal{O}(z_{12}^u,z_2) - \mathcal{O}(z_1,z_{21}^u)\Big]\,,
\end{eqnarray}
where $\mathbb{P}_{12}$ is the permutation operator
\begin{align}
  \mathbb{P}_{12} \mathcal{O}(z_1,z_2) &= \mathcal{O}(z_2,z_1)\,.
\end{align}
One obtains (only the non-vanishing contributions are listed):
\begin{eqnarray}
%%%% diagram (a)
  h_{(a)}(u) &=& C_F^2\frac{\bar u}{u}\left[\ln u+1\right],
\nonumber\\
%%%% diagram (b)
  h_{(b)}(u) &=&
 C_F \frac{\bar u}{u}
 \Big[ (2C_A-\beta_0 ) \ln \bar u + \frac83 C_A  - \frac53 \beta_0 \Big],
%\hspace*{2cm} \beta_0 = \frac{11}{3}C_A - \frac23 n_f,
\nonumber\\
%%%% diagram (c)
  h_{(c)}(u) &=&
\Big[C_F^2 - \frac12 C_FC_A\Big] \frac{\bar u}{u} \Big[\ln^2\bar u - 3 \frac{u}{\bar u}\ln u  + 3 \ln \bar u -\ln u - 1\Big],
\nonumber\\
%%%% diagram (d)
  h_{(d)}(u) &=& \frac12 C_F C_A
\frac{\bar u}{u} \left[ \frac12 \left(1-\frac{u}{\bar u}\right)\ln^2 u +\ln \bar u -3\right],
\nonumber\\
%%%% diagram (e)+(f)
  \hspace{-2mm}h_{(e+f)}(u) &=&
   2\, C_F^2 \frac{\bar u}{u} \left[ 2\big( \Li_2(1)-\Li_2(\bar u)\big) - \ln^2\bar u + 2 \frac{u}{\bar u}\ln u\right]
\nonumber\\
    &&{}+ C_F C_A  \frac{\bar u}{u} \biggl[2 \big(\Li_2(\bar u)- \Li_2(u)\big) + \frac12 \ln^2\bar u
     - \frac12 \ln^2 u - \frac{1+ u}{\bar u}\ln u-2 \biggr],
\nonumber\\
%%%% diagram (g)
  h_{(g)}(u) &=& -  C_F C_A \frac{\bar u}{u}
\biggl[
   \Li_2(\bar u)-\Li_2(1)+1+\frac{1}{4} \ln^2\bar u
+\ln\bar u -\frac{1+u}{2\bar u}  \ln u \left(\frac12 \ln u +1\right)\biggr],
\nonumber\\
%%%% diagram (j)
 h_{(j)}(u) &=&\Big[C_F^2 - \frac12 C_FC_A\Big] \,\ln u\,,
\nonumber\\
%%%% diagram (o)
  h_{(o)}(u) &=&2\Big[C_F^2 - \frac12 C_FC_A\Big]\frac{\bar u}{u}
\biggl[-2\Li_2(u)+\frac{u}{\bar u}\ln u\ln\bar u-\frac12 \ln^2 \bar u-\frac{u}{\bar u}\ln u\biggr],
\nonumber\\
%%%% diagram (p)
 h_{(p)}(u) &=&C_FC_A\frac{\bar u}u
 \biggl[\Li_2( u)+\frac1{\bar u}\ln u\ln\bar u-\frac14 \ln^2\bar u-\frac{u}{4\bar u}\ln^2u-\frac{u}{\bar u}\ln u\biggr],
\end{eqnarray}
and
\begin{align}
%diagram (h)
\chi_{(h)}(\alpha,\beta)&=-C_F^2\Big[\ln\alpha+3 \Big],\,
\notag\\
%diagram (i)
\chi_{(i)}(\alpha,\beta)&=C_F\biggl[\frac16(C_A-\beta_0)\delta(\alpha)\delta(\beta)-\left(C_A-\frac12\beta_0\right)\ln(1-\alpha-\beta)
-\frac{10}3 C_A+\frac{11}6\beta_0\biggr],
\notag\\
%diagram (j)
\chi_{(j)}(\alpha,\beta)&=\Big[C_F^2 - \frac12 C_FC_A\Big]
%\biggl[\ln^2\bar\alpha- \ln^2(1-\alpha-\beta)- 4\ln(1-\alpha-\beta)-3\ln\bar\tau - \frac12\ln\tau -6+\delta(\alpha)\delta(\beta)\biggr],
\biggl[\ln^2\bar\alpha -8\ln\bar\alpha- \ln^2(1-\alpha-\beta)-7\ln\bar\tau  - \frac12\ln\tau -6+\delta(\alpha)\delta(\beta)\biggr],
\notag\\
%diagram (k)
\chi_{(k)}(\alpha,\beta)& =-\frac12 C_FC_A\Big[\ln(1-\tau)+2\ln\tau-4+\ln^2\alpha-\ln^2\bar\alpha\Big],
\notag\\
%diagram (l)
\chi_{(l)}(\alpha,\beta)& = C_F^2\left[2\ln\bar \tau + 1 + \frac12\ln^2(1-\alpha-\beta) - \ln^2\bar\alpha\right],
\notag\\
%diagram (m)
\chi_{(m)}(\alpha,\beta)& = \Big[C_F^2 - \frac12 C_FC_A\Big]\Big[\ln^2\bar\alpha+4\ln\bar\alpha\Big],
\notag\\
%diagram (n)
\chi_{(n)}(\alpha,\beta)& =C_F^2\Big[2\ln\tau+8+4\big(\Li_2(\alpha)-\Li_2(1)\big)+\ln^2\alpha+\ln^2\bar\alpha+2\ln\alpha \Big],
\notag\\
%diagram (o)
\chi_{(o)}(\alpha,\beta)& = \Big[C_F^2 - \frac12 C_FC_A\Big] \biggl[%5\ln\bar \tau
2\left(2+\frac{1}{\tau}\right)\ln\bar\tau-3\ln\tau -\ln^2\bar\tau-2\Li_2(\tau)+\ln^2(1-\alpha-\beta)
\notag\\
&\quad % + \ln(1-\alpha-\beta)
%+\frac{2\alpha}{\bar\alpha}\ln\alpha
- \ln^2(\alpha\bar\alpha) - 4\Big[\Li_2(\alpha)-\Li_2(1)\Big]
+ \frac2{\alpha}\ln\bar\alpha - 2\big[2+\Li_2(1)-3\zeta(3)\big] \delta(\alpha)\delta(\beta)\biggr],
%-\frac{2}{\bar\alpha}\ln\alpha\biggr]
%-2\ln\alpha
\notag\\
%diagram (p)
\chi_{(p)}(\alpha,\beta)& =  C_F C_A\left[-\frac34\ln\tau + \Li_2(\bar\alpha) - \Li_2(\alpha) +\frac1\alpha\ln\bar\alpha +
 \big[\Li_2(1)-2\big]\delta(\alpha)\delta(\beta)\right].
\end{align}
The nonvanishing contributions to $\chi^{\mathbb{P}}(\alpha,\beta)$ originate from two diagrams only:
\begin{align}
%diagram (m)
\chi_{(m)}^{\mathbb{P}}(\alpha,\beta) &=-\Big[C_F^2 - \frac12 C_FC_A\Big]\left[4\ln\bar\tau-2\ln\bar\alpha^2+\ln^2(1-\alpha-\beta)\right],
\notag\\
%diagram (o)
\chi_{(o)}^{\mathbb{P}}(\alpha,\beta) &=\Big[C_F^2 - \frac12 C_FC_A\Big]
\Big[6\ln\bar\tau%-\ln\left(1-\frac{\bar\beta\bar w}{\beta w}\right)
-\ln^2\bar\tau-2\bar\tau\ln\bar\tau
-2\ln^2\bar \alpha+\ln^2(1-\alpha-\beta)\Big].
\end{align}
In all expressions here and below
\begin{align}
   \tau = \frac{\alpha\beta}{\bar\alpha\bar\beta}\,, &&  \beta_0 = \frac{11}{3}C_A - \frac23 n_f\,.
\end{align}

%\newpage

%%%%%%%%%%%%%%%%%%%%%%%%%%%%%%%%%%%%%%%%%%%%%%%%%%%%%%%%%%%%%%%%%%%%%%%%%%%%%%%%%%%%%%%%%%
\subsection{Conformal anomaly}
%%%%%%%%%%%%%%%%%%%%%%%%%%%%%%%%%%%%%%%%%%%%%%%%%%%%%%%%%%%%%%%%%%%%%%%%%%%%%%%%%%%%%%%%%%

Terms due to the conformal variation of the action can be written in the form
\begin{eqnarray}
\Delta S_+  &=& \frac12 \mathbb{H}(z_1+z_2) + z_{12} \Delta_+
\end{eqnarray}
where $\mathbb{H}$ is the corresponding contribution to the evolution kernel.
The contributions to $\Delta_+$ from the diagrams in Fig.~\ref{fig:Diagrams}  %(a)--(g)
(including symmetric diagrams with the interchange of the quark and the antiquark)
can be brought to the following form:
\begin{eqnarray}
  [\Delta_+ \mathcal{O}](z_1,z_2) &=&
\int_0^1\!d\alpha\!\int_0^{\bar\alpha} d\beta \Big[\omega(\alpha,\beta) + \omega^{\mathbb{P}}(\alpha,\beta) \mathbb{P}_{12}\Big]
\Big[\mathcal{O}(z_{12}^{\alpha},z_{21}^\beta)-\mathcal{O}(z_{12}^{\beta},z_{21}^\alpha)\Big]
\nonumber\\
&&{}
+ \int_0^1\!du\int_0^1\!dt \, \varkappa(t)\,\Big[\mathcal{O}(z_{12}^{ut},z_2) - \mathcal{O}(z_1,  z_{21}^{ut})\Big].
\end{eqnarray}
We obtain (only nonvanishing contributions are listed)
\begin{eqnarray}
%%%% diagram (a)
  \varkappa_{(a)}(t) &=& C_F^2\left[\frac1t+\frac{1+\bar t}{t}\ln t\right],
\nonumber\\
%%%% diagram (b)
  \varkappa_{(b)}(t) &=&  - 2 C_F \frac{\bar t}{t}
 \Big[ (\beta_0 -2 C_A ) \ln \bar t -  \frac83 C_A  + \frac53 \beta_0 \Big],
\nonumber\\
%%%% diagram (c)
\varkappa_{(c)}(t) &=& \Big[C_F^2 - \frac12 C_FC_A\Big]\Big[t \ln ^2 t  + \frac{2\bar t}{t}\ln ^2\bar t  + \frac{6\bar t}{t} \ln \bar t
- \frac{\bar t}{t} (3t+2) \ln t  - 9 t + 8 -\frac{1}{t}\Big],
\nonumber\\
%%%% diagram (d)
\varkappa_{(d)}(t) &=& C_F C_A\biggl\{ \frac{\bar t}{t} \left[ %\frac12 % \left(1-\frac{t}{\bar t}\right)
\frac{1-2t}{2\bar t}\ln^2 t +\ln \bar t -3\right]
+\frac12\biggl[ \frac12\ln^2 t-\bar t\ln^2\bar t+\frac{t^2-\bar t}{t}\ln t-2\bar t\ln\bar t-1-\bar t\biggr]
%-\frac12\Big[ \bar t  \ln^2\bar t+ 2 \bar t  \ln\bar t+2 \bar t \ln t - \ln t + t \Big], \nonumber\\&& -\frac12\Big[
%-\frac{1}{2} \ln^2 t+\frac{\ln t}{t}-2\ln t  - \bar t \ln\bar t +2\bar t \Big] -\frac12\Big[ \bar t \ln\bar t + t
%\ln t \Big]
\biggr\},
\nonumber\\
%%%% diagram (e+f)
\varkappa_{(e+f)}(t) &=&
 -4C_F^2\bigg\{ t\Big[\Li_2(t)-\Li_2(1)\Big] + 2 \frac{\bar t}{t} \Big[\Li_2(\bar t)-\Li_2(1)\Big]
 + \frac{\bar t}{t} \ln^2 \bar t + \frac12 t \ln^2 t + 2 \bar t \ln\bar t
\nonumber\\&& - \frac32(1-2t)\ln t + 2 \biggr\}
+ C_F C_A \frac{\bar t}{t} \biggl\{
4 \Big[\Li_2(\bar t)-\Li_2(t)\Big] + \frac12(2+t)\ln^2\bar t
\nonumber\\&&{}
- \left(1-\frac{t^2}{2\bar t}\right)\ln^2 t
- 2(1-2t)\ln\bar t - \left(5t+ \frac{1}{\bar t}\right)\ln t - 3 + 2t\biggr\},
\nonumber\\
%%%% diagram (g)
\varkappa_{(g)}(t) &=& C_F C_A \frac{\bar t}{t}\biggl\{ t\Big[\Li_2(\bar t) - \Li_2(1)\Big]+ \frac14 t \ln^2\bar t + \frac14(2\!+\!t)\ln^2 t -
           (3\!-\!t)\ln\bar t
\nonumber\\&&{}
+ \frac12 \left(1-\frac{t^2}{\bar t}\right)\ln t  -\bar t -\frac32\biggr\},
\nonumber\\
%%%% diagram (j)
  \varkappa_{(j)}(t) &=& \Big[C_F^2 - \frac12 C_FC_A\Big]\Big[-t\ln t -1\Big],
\nonumber\\
%%%% diagram (o)
  \varkappa_{(o)}(t) &=& \Big[C_F^2 - \frac12 C_FC_A\Big]
\Big\{
\frac4{\bar t}\Big[\Li_2(t)-\Li_2(1)\Big]-4t\Big[\Li_2(t)-\Li_2(1)\Big]+4\bar t\Li_2(1)
\notag\\&&{}
-2t\ln t\ln\bar t +\frac t{\bar t}\ln^2t+\bar t
\ln^2\bar t-4t\ln\bar t+\frac{2t}{\bar t}(2-3t)\ln t+2
\Big\},
\nonumber\\
%%%% diagram (p)
  \varkappa_{(p)}(t) &=& C_F C_A \Big\{
\frac{2t}{\bar t}\Big[\Li_2(t)-\Li_2(1)\Big]+\bar t \Big[\Li_2(\bar t)-\Li_2(1)\Big]
-t\ln t\ln\bar t
\nonumber\\&&{}
+\frac14\bar t\ln^2\bar t
+\frac14\frac{t(3-t)}{\bar t}\ln^2t
-\frac{t^2}{\bar t}\ln t+\frac12\ln t-\frac{1+t}{t}\ln\bar t + 1\Big\}.
\end{eqnarray}
 The function $\omega^{\mathbb{P}}(\alpha,\beta)$ originates from two diagrams only:
\begin{align}
%%%% diagram (m)
  \omega^{\mathbb{P}}_{(m)}(\alpha,\beta) &= -{C_F}\left(C_F-\frac12 C_A\right)\, 2\beta\Big(\ln^2\bar\alpha+4\ln\bar\alpha\Big),
\nonumber\\
%%%% diagram (o)
  \omega^{\mathbb{P}}_{(o)}(\alpha,\beta) &=
  -4 {C_F}\left(C_F-\frac12 C_A\right)\,\Biggl\{
\left(\bar\alpha-\frac1{\bar\alpha}\right)\Big[\Li_2\left(\frac\alpha{\bar\beta}\right)-\Li_2(\alpha)-\ln\bar\alpha\ln\bar\beta\Big]
+\alpha \bar\tau \ln\bar\tau
\notag\\
&\quad
+\frac{\beta^2}{\bar\beta}\ln\bar\alpha -\frac12 \beta\Big(\ln^2\bar\alpha+4\ln\bar\alpha\Big)\Biggr\}.
\end{align}
%\begin{eqnarray}
%  [\Delta \mathcal{O}]^{(a-g)} &=&
%\int_0^1\!du\int_0^1\!dt \, q(t)\Big[\mathcal{O}(z_{12}^{ut},z_2) - \mathcal{O}(z_1,  z_{21}^{ut})\Big]
%\end{eqnarray}
The nonvanishing contributions to $\omega(\alpha,\beta)$ are
\begin{eqnarray}
%%%% diagram (h)
  \omega_{(h)}(\alpha,\beta) &=& C_F^2\,
\bar\beta\Big[\ln\alpha+3 \Big],
\nonumber\\[2mm]
%%%%% diagram (i)
%  \omega_{(i)}(\alpha,\beta) &=& 0
%\nonumber\\[2mm]
%%%% diagram (j)
  \omega_{(j)}(\alpha,\beta) &=& \Big[C_F^2 - \frac12 C_FC_A\Big]
\Big[\alpha\ln^2\alpha+\beta\ln^2\bar\alpha +7\beta\ln\bar\alpha
+\bar\beta\ln\alpha +4\alpha\ln\alpha-4 \alpha\Big],
\nonumber\\[2mm]
%%%% diagram (k)
  \omega_{(k)}(\alpha,\beta) &=& C_F C_A
\Big[\frac12\beta\ln^2\bar\alpha+\frac12\bar\beta \ln^2\alpha
+ 3\beta\ln\bar\alpha + 2 \bar\beta \ln\alpha+\frac12\alpha\ln\alpha-\frac12\bar\alpha\ln\bar\alpha-\alpha\Big],
\nonumber\\[2mm]
%%%% diagram (l)
  \omega_{(l)}(\alpha,\beta) &=& - C_F^2\,\beta
\Big[\ln^2\bar\alpha+4\ln\bar\alpha -2\Big],
\nonumber\\[2mm]
%%%% diagram (m)
  \omega_{(m)}(\alpha,\beta) &=& \Big[C_F^2 - \frac12 C_FC_A\Big]
\Big[  \bar\alpha\ln\bar\alpha\big(\ln\bar\alpha+2\big)
+\beta\big(\ln^2\bar\alpha+4\ln\bar\alpha -2\big)
 \Big],
\nonumber\\[2mm]
%%%% diagram (n)
  \omega_{(n)}(\alpha,\beta) &=& 2 C_F^2
\biggl\{(\beta-\alpha)\left[4+3\ln\alpha-2\ln\bar\alpha+\frac12\ln^2\alpha+\frac12\ln^2\bar\alpha+2\big(\Li_2(\alpha)-\Li_2(1)\big)\right]
\nonumber\\
&&{}-\frac2{\bar\alpha}\Big[\Li_2(\alpha)-\Li_2(1)\Big]-\frac{1}{2\bar\alpha}\ln^2\alpha
+\frac12\ln^2\bar\alpha
+\Big[\frac{\alpha}{\bar\alpha}-3+\alpha \Big]\ln\alpha
%\nonumber\\
%&&{}
+\bar\alpha\ln\bar\alpha +\alpha
\biggr\},
\nonumber\\
%%%% diagram (o)
  \omega_{(o)}(\alpha,\beta) &=& -4\Big[C_F^2 - \frac12 C_FC_A\Big]
 \biggl\{\left(\beta-\frac1\beta\right)\Big[\Li_2(\alpha/\bar\beta)-\Li_2(\alpha)-\Li_2(\beta)\Big]
\nonumber\\&&{}
+\Big(\beta-\frac1{\bar\alpha}\Big)\Big[\Li_2(\alpha)-\Li_2(1)\Big]
+\frac12(\alpha+\beta)\ln\alpha\ln\bar\alpha
+\frac12\beta\ln^2\bar\alpha
\nonumber\\&&{}
- \frac14\frac{\bar\alpha}\alpha\ln^2\bar\alpha-\frac14 \frac\alpha{\bar\alpha}\ln^2\alpha
+ \alpha\frac{\bar\tau}{\tau}\ln\bar\tau + 3 \beta\ln\bar\alpha
+\frac{\alpha}{\bar\alpha}\ln\alpha
-\frac12 \frac{\bar\alpha}\alpha\ln\bar\alpha
\nonumber\\&&{}
- (1-\alpha-\beta)\Big[
\frac14\ln^2\alpha-\frac14\ln^2\bar\alpha
+\frac32\ln\alpha-\frac32\ln\bar\alpha
-\frac1{2\alpha}\ln\bar \alpha
\Big]
\biggr\},
\nonumber\\
%%%% diagram (p)
  \omega_{(p)}(\alpha,\beta) &=& C_FC_A
\biggl\{
\frac{2}{\bar\alpha} \Big[\Li_2(\alpha)-\Li_2(1)\Big]+\frac{2}{\alpha} \Big[\Li_2(\bar \alpha)-\Li_2(1)\Big]
+\frac12\frac{\bar\alpha}\alpha\ln^2\bar\alpha
\nonumber\\&&{}
+\frac12\frac\alpha{\bar\alpha}\ln^2\alpha
-\frac2{\bar\alpha}\ln\alpha-\frac2\alpha\ln\bar\alpha
+\bar\alpha\ln \alpha+\alpha\ln\bar\alpha+\frac32\ln\bar\alpha+\frac12\ln\alpha
\nonumber\\&&{}
+ \frac32\alpha\ln\tau
-(\alpha-\beta) \Big[
\frac1\alpha\ln\bar\alpha
+\Li_2(\bar\alpha)-\Li_2(\alpha)
\Big]\biggr\}.
\end{eqnarray}
Note that the only contribution of the diagram in Fig.~\ref{fig:Diagrams}(i) is through the corresponding
term $\sim \chi_{(i)}(\alpha,\beta)$ in the evolution kernel.

%
%%%%%%%%%%%%%%%%%%%%%%%%%%%%%%%%%%%%%%%%%%%%%%%%%%%%%%%%%%%%%%%%%%%%%%%%%
%%%%%%%%%%%%%%%%%%%%%%%%%%%%%%%%%%%%%%%%%%%%%%%%%%%%%%%%%%%%%%%%%%%%%%%%%
%

\end{document}